\documentclass[longauth]{aa}
\pdfoutput=1                            
\usepackage[varg]{txfonts}

\usepackage{amsmath}
\usepackage{amsfonts}
\usepackage{amssymb}
\usepackage{wasysym}
\usepackage{graphicx}

\usepackage{natbib}
\usepackage{epsfig}
\usepackage{xspace}
\usepackage{booktabs}
\usepackage{dcolumn}
\usepackage[para]{footmisc}
\usepackage{placeins}
\usepackage{multicol}

\usepackage[switch]{lineno}

\usepackage[utf8]{inputenc}
\usepackage[english]{babel}

\def\ms{\,m\,s$^{-1}$}         %m.s -1
\def\kms{\,km\,s$^{-1}$}         %m.s -1
\def\ms{\hbox{\,m\,s$^{-1}$}}         %m.s -1
\def\m2s2{\hbox{\,m$^{2}$\,s$^{-2}$}} %m2.s -2
\def\kms{\hbox{\,km\,s$^{-1}$}}       %km.s -1
\def\Msun{\hbox{$M_{\odot}$}}             %Msun
\def\Rsun{\hbox{$R_{\odot}$}}

\newcommand{\emth}[1]{\ensuremath{#1}\xspace}
\newcommand{\parm}[1]{\ensuremath{#1}\xspace}

\newcommand{\gcm}{\emth{\mathrm{g\,cm^{-3}}}}
\newcommand{\mjup}{\emth{\mathrm{M_{Jup}}}} 	% Jupiter mass
\newcommand{\rjup}{\emth{\mathrm{R_{Jup}}}} 	% Jupiter radius
\newcommand{\mearth}{\emth{\mathrm{M_{\oplus}}}} 	% Jupiter mass

\newcommand{\vsini}{\emth{v\sin{i}}\xspace}

\newcommand{\teff}{T$_{\rm eff}$}
\newcommand{\logg}{log {\it g}}

\newcommand{\vmicro}{$v_{\rm micro}$}
\newcommand{\vmacro}{$v_{\rm macro}$}

\newcommand{\iid}{i.i.d.\xspace}

\newcommand{\pvec}{\ensuremath{\vec{\theta}}\xspace}
\newcommand{\Dlc}{\ensuremath{\vec{D_{\mathrm{LC}}}}\xspace}
\newcommand{\Drv}{\ensuremath{\vec{D_{\mathrm{RV}}}}\xspace}

\newcommand{\pfr}{\parm{f}}           		% Flux ratio
\newcommand{\prr}{\parm{k}}           		% Radius ratio
\newcommand{\pper}{\parm{p}}          		% Period
\newcommand{\pec}{\parm{e}}           		% Eccentricity
\newcommand{\pom}{\parm{\omega}}      		% Argument of periastron
\newcommand{\pimp}{\parm{b}}           		% Inclination

\newcommand{\pelc}{\parm{\sigma_\mathrm{l}}} 		% Long cadence error
\newcommand{\pelt}{\parm{\sigma_\mathrm{l,t}}} 		% Long cadence error
\newcommand{\pelr}{\parm{r_\mathrm{l}}} 		% Long cadence error
\newcommand{\pesc}{\parm{\sigma_\mathrm{s}}} 		% Short cadence error
\newcommand{\pest}{\parm{\sigma_\mathrm{s,t}}} 		% Short cadence error
\newcommand{\pesr}{\parm{r_\mathrm{s}}} 		% Short cadence error
\newcommand{\pcf}{\parm{c}}       		% Contamination
\newcommand{\ptc}{\emth{T_\mathrm{0}}} 		% Eclipse center
\newcommand{\prtd}{\emth{2/T_1}} 		% Transit center
\newcommand{\prvz}{\emth{C}} 		% RV zeropoint
\newcommand{\prva}{\emth{K}} 		% RV zeropoint

\newcommand{\smass}{\emth{\mathrm{M_\star}}}
\newcommand{\srad}{\emth{\mathrm{R_\star}}}
\newcommand{\pmass}{\emth{\mathrm{M_P}}}	% Planet mass
\newcommand{\lcad}{c$_\mathrm{l}$\xspace}
\newcommand{\scad}{c$_\mathrm{s}$\xspace}

\begin{document}
\title{Transiting exoplanets from the CoRoT space mission\thanks{The CoRoT space mission, launched on December 27, 2006, has been developed and is operated by CNES, with the contribution of Austria, Belgium, Brazil , ESA (RSSD and Science Programme), Germany, and Spain.}}
\subtitle{XXV. CoRoT-27b: a massive and dense planet on a short-period orbit}

\author{
%%%% COROT EXO COIs
 Parviainen, H.\inst{\ref{iiac},\ref{iull},\ref{ioxford}}
\and Gandolfi, D.\inst{\ref{iinaf}} 
\and Deleuil, M.\inst{\ref{ilam}} 
\and Moutou, C.\inst{\ref{ilam}} 
\and Deeg,~H.~J.\inst{\ref{iiac},\ref{iull}} 
\and Ferraz-Mello, S.\inst{\ref{iiag}}
\and Samuel,~B.\inst{\ref{iparis}}
\and Csizmadia, Sz.\inst{\ref{iberlin}} 
\and Pasternacki, T.\inst{\ref{iberlin}} 
\and Wuchterl,~G.\inst{\ref{itl}}
\and Havel,~M.\inst{\ref{iames}}
\and Fridlund,~M.\inst{\ref{iesa}}
\and Angus,~R.\inst{\ref{ioxford}}
\and Tingley,~B.\inst{\ref{iaarhus}} 
\and Grziwa,~S.\inst{\ref{iriuu}}
\and Korth,~J.\inst{\ref{iriuu}}
\and Aigrain,~S.\inst{\ref{ioxford}} 
\and Almenara,~J.~M.\inst{\ref{ilam}}
\and Alonso,~R.\inst{\ref{iiac},\ref{iull}}
% \and Auvergne, M.\inst{\ref{iparis}} 
\and Baglin, A.\inst{\ref{iparis}}
\and Barros,~S.C.C.\inst{\ref{ilam}}
% \and Bonomo,~A.~S.\inst{\ref{ilam},\ref{iinaf}}
\and Bord\'e,~P.\inst{\ref{iias}} % 
\and Bouchy,~F.\inst{\ref{iohp},\ref{iiap}} %
\and Cabrera, J.\inst{\ref{iberlin}} 
% \and Damiani,~C.\inst{\ref{ilam}}
\and D\'iaz,~R.~F.\inst{\ref{ilam}}
\and Dvorak,~R.\inst{\ref{ivienna}} % 
\and Erikson, A.\inst{\ref{iberlin}}
% \and Gillon,~M.\inst{\ref{igeneve}} 
\and Guillot,~T.\inst{\ref{ioca}} 
\and Hatzes, A.\inst{\ref{itl}} 
\and H\'ebrard,~G.\inst{\ref{iiap},\ref{iohp}}
% \and Llebaria, A.\inst{\ref{ilam}} % 
% \and Lammer, H.\inst{\ref{isri}} 
\and Mazeh, T.\inst{\ref{ispa}} 
\and Montagnier,~G.\inst{\ref{iiap},\ref{iohp}} 
\and Ofir,~A.\inst{\ref{igot}}
\and Ollivier,~M.\inst{\ref{iias}} % 
\and P\"atzold, M.\inst{\ref{iriuu}} 
% \and Queloz, D.\inst{\ref{igeneve}}
\and Rauer, H.\inst{\ref{iberlin}} 
\and Rouan,~D.\inst{\ref{iparis}}
\and Santerne,~A.\inst{\ref{iporto}}
\and Schneider,~J.\inst{\ref{iluth}} 
%%% + COROT ASSOCIATED SCIENTISTS
%%% and please ask the CoIs to identify who in their team is missing and has contributed significantly to the discovery.
% \and H\'ebrard, G.\inst{6} 
% \and Guenther, E.\inst{14}
% \and Carpano, S.\inst{11}
% \and Carone, L.\inst{17} 
}

\institute{
Instituto de Astrof\'isica de Canarias (IAC), E-38200 La Laguna, Tenerife, Spain\label{iiac}
\and Dept. Astrof\'isica, Universidad de La Laguna (ULL), E-38206 La Laguna, Tenerife, Spain\label{iull}
\and Department of Physics, Denys Wilkinson Building Keble Road, Oxford, OX1 3RH\label{ioxford}
\and INAF-Catania Astrophysical Observatory, Via S. Sofia 78, I-95123 Catania, Italy \label{iinaf}
\and Aix Marseille Universit\'e, CNRS, LAM (Laboratoire d'Astrophysique de Marseille) UMR 7326, 13388, Marseille, France \label{ilam}
\and IAG, Universidade de Sao Paulo, Brazil\label{iiag}
\and LESIA, UMR 8109 CNRS , Observatoire de Paris, UVSQ, Universit\'e Paris-Diderot, 5 place J. Janssen, 92195 Meudon cedex, France\label{iparis}
\and Institute of Planetary Research, German Aerospace Center, Rutherfordstrasse 2, 12489 Berlin, Germany\label{iberlin}
\and Th\"uringer Landessternwarte, CoRoT (DLR), Sternwarte 5, Tautenburg, D-07778 Tautenburg, Germany\label{itl}
\and NASA Ames Research Center, MS 244-30, P.O. Box 1, 94035-0001 Moffett Field, USA\label{iames}
\and Research and Scientific Support Department, ESTEC/ESA, PO Box 299, 2200 AG Noordwijk, The Netherlands\label{iesa}
\and Department of Physics and Astronomy, Aarhus University, Ny Munkegade 120, 8000 Aarhus C, Denmark\label{iaarhus}
\and Rheinisches Institut f\"ur Umweltforschung an der Universit\"at zu K\"oln, Aachener Strasse 209, 50931, Germany\label{iriuu}
\and Institut d'Astrophysique Spatiale, Universit\'e Paris XI, F-91405 Orsay, France\label{iias}
\and Observatoire de Haute Provence, 04670 Saint Michel l'Observatoire, France\label{iohp}
\and Institut d'Astrophysique de Paris, 98bis boulevard Arago, 75014 Paris, France\label{iiap}
\and University of Vienna, Institute of Astronomy, T\"urkenschanzstr. 17, A-1180 Vienna, Austria\label{ivienna}
\and Observatoire de la C\^ote d'Azur, Laboratoire Cassiop\'ee, BP 4229, 06304 Nice Cedex 4, France\label{ioca}
\and School of Physics and Astronomy, Raymond and Beverly Sackler Faculty of Exact Sciences, Tel Aviv University, Tel Aviv, Israel\label{ispa}
\and Institut f\"ur Astrophysik, Georg-August-Universit\"at, Friedrich-Hund-Platz 1, 37077 G\"ottingen, Germany\label{igot}
\and Centro de Astrof\'isica, Universidade do Porto, Rua das Estrelas, 4150-762 Porto, Portugal\label{iporto}
\and LUTH, Observatoire de Paris, CNRS, Universit\'e Paris Diderot; 5 place Jules Janssen, 92195 Meudon, France\label{iluth}
% \and Observatoire de l'Universit\'e de Gen\`eve, 51 chemin des Maillettes, 1290 Sauverny, Switzerland\label{igeneve}
% \and Space Research Institute, Austrian Academy of Science, Schmiedlstr. 6, A-8042 Graz, Austria\label{isri}
% \and University of Li\`ege, All\'ee du 6 ao\^ut 17, Sart Tilman, Li\`ege 1, Belgium
% \and INAF - Osservatorio Astronomico di Torino, via Osservatorio 20, 10025, Pino Torinese, Italy\label{iinaf}
}
\date{Received ; accepted }

\abstract{}
  {We report the discovery of a massive and dense transiting planet CoRoT-27b on a 3.58-day orbit around a 4.2~Gyr-old G2~star. The planet candidate was identified from the CoRoT photometry, and was confirmed as a planet with ground-based spectroscopy.}
  {The confirmation of the planet candidate is based on radial velocity observations combined with imaging to rule out blends. The characterisation of the planet and its host star was carried out using a Bayesian approach where all the data (CoRoT photometry, radial velocities, and spectroscopic characterisation of the star) are used jointly. The Bayesian analysis included a study whether the assumption of white normally distributed noise holds for the CoRoT photometry and whether the use of a non-normal noise distribution offers advantages in parameter estimation and model selection.}
  {CoRoT-27b has a mass of $10.39 \pm 0.55$~\mjup, a radius of $1.01 \pm 0.04$~\rjup, a mean density of $12.6_{-1.67}^{+1.92}$~\gcm, and an effective temperature of $1500 \pm 130$~K. The planet orbits around its host star, a 4.2~Gyr-old G2-star with a mass $\smass=1.06$~\Msun and a radius $\srad=1.05$~\Rsun, on a $0.048 \pm 0.007$~AU orbit of 3.58~days. The radial velocity observations allow us to exclude highly eccentric orbits, namely, $e<0.065$ with 99\% confidence. Given its high mass and density, theoretical modelling of CoRoT-27b is demanding. We identify two solutions with heavy element mass fractions of $0.11\pm0.08$~\mearth and $0.07\pm0.06$~\mearth, but even solutions void of heavy elements cannot be excluded.
  We carry out a secondary eclipse search from the CoRoT photometry using a method based on Bayesian model selection, but conclude that the noise level is too high to detect eclipses shallower than a 9\% the transit depth.
  }
  {}


\titlerunning{}
\authorrunning{}

\maketitle

% \linenumbers

\section{Introduction}
\label{sec:introduction}
We report the discovery of a new massive high-density transiting planet on a short-period orbit, CoRoT-27b. The planet falls within the scarcely populated overlapping mass regime between planets and brown dwarfs \citep{Leconte2009,Baraffe2010}, and contributes to our understanding of the high-mass tail of the planet population.

Distinguishing between high-mass planets and low-mass brown dwarfs is an ambiguous task that depends on the definition of a planet \citep{Schneider2011}. If we decide to use the formation history as the discriminating factor---naming objects formed by core accretion as planets, and objects formed by gravitational collapse as brown dwarfs---we may be able to identify sets of observables characteristic to the two populations. 
Thus, while a mass estimate is not enough to distinguish between the formation histories, the differences in the observable distributions (orbital eccentricity, host rotation rate, host metallicity,  etc.) may allow us to infer the likely formation history for a given object based on a probabilistic model.

The currently known population of massive short-period planets has shown some tentative trends towards higher orbital eccentricities and host star rotation rates and lower host star metallicities than observed for the lower-mass planets \citep{Southworth2009c,Bakos2011b}.

Furthermore, the massive close-in planets are predominantly found around F-type stars \citep{Bouchy2011}, Kepler-75b~\citep{Hebrard2013} being the only massive short-period star found around a G-type star before CoRoT-27b. \citet{Bouchy2011} propose that this trend is real instead of an observation bias and that it is due to differences in tidal braking by G- and F-dwarfs. The massive close-in companions around G-dwarfs would rapidly migrate inwards and be engulfed by the host star owing to the star's strong tidal braking, while the companions around F-dwarfs with weaker tidal braking would be spared this fate \citep{Barker2009}.

While the trends in eccentricity and host star properties are based on small number statistics, already the large variability in these properties may be a sign of mixing objects of different nature. While the detection of possible distinguishable populations will require a proper cluster analysis with significantly more objects, each new planet discovery in this mass and period regime will get us a bit closer to understanding the fringe regions of the planet and brown dwarf distributions.

\section{Data} 
\label{sec:data}
\subsection{CoRoT light curve}
The CoRoT satellite offers two time cadences. The survey mode delivers data with a cadence of 512~s (long cadence, \lcad, from here on), created by stacking 16 exposures of 32~s. For the planet candidates identified during an observing run, a fast time sampling of 32~s (short cadence, \scad) is also available \citep{Surace2008}. Furthermore, data of bright ($\lesssim R=13.5$mag) targets have been acquired with three-colour photometry, while the fainter ones are only observed in a single passband. 

CoRoT-27 was observed continuously for 83.5 days (from 8~July~2011 to 30~September~2011) in the monochromatic mode during the LRc08 run towards the galactic centre. We present the catalogue IDs, coordinates, and magnitudes in Table~\ref{table:star}.
The star was first observed in the \lcad~mode, and the mode was changed to \scad after a promising transit candidate with a depth of $\sim 1\%$ was discovered. 

\begin{table}[t]
\caption{ CoRoT-27 IDs, coordinates, and magnitudes.}            
\begin{minipage}[!]{\columnwidth}  
\renewcommand{\footnoterule}{}     
\begin{tabular*}{\columnwidth}{@{\extracolsep{\fill}} lrc}
\toprule\toprule               
CoRoT window ID & LRc08\_E2\_4905  \\
CoRoT ID & 652180928        \\
USNO-A2 ID  & 0900-13156792 \\
2MASS ID   &  1183241962    \\
\\
\multicolumn{2}{l}{Coordinates} \\
\midrule            
RA (J2000)  &  18:33:59  \\
Dec (J2000) &  +5:32:18.503  \\
\\
\multicolumn{3}{l}{Magnitudes} \\
\midrule
\centering
Filter & Mag & Error \\
\midrule
B$^a$  & 16.502 & \\
V$^a$  & 15.540 & \\
r'$^a$ & 15.848 & \\
% i'$^a$ & & \\
J$^b$  & 13.571 & 0.024 \\
H$^b$  & 13.170 & 0.031 \\
K$^b$  & 12.985 & 0.032 \\
\bottomrule
\vspace{-0.5cm}
\footnotetext[1]{Provided by Exo-Dat \citep{Deleuil2009};}
\footnotetext[2]{from 2MASS catalogue.}
\end{tabular*}
\end{minipage}
\label{table:star}      
\end{table}

The light curve consists of 6920 \lcad and 114368 \scad points in total, of which 788 and 5897 are marked as extreme outliers by the CoRoT pipeline.
Further 1022 and 9869 points are marked as exposures obtained during the crossing of South Atlantic Anomaly (SAA), a region of the satellite's orbit where it is exposed to high doses of radiation~\citep{PinheirodaSilva2008}. We exclude the points marked as extreme outliers from the analysis, but keep the inside-SAA points. 
The rationale behind keeping the inside-SAA points is that while the number of high-energy particle hits increases during the SAA crossings, the true number of points affected by the particle-hit events does not justify the removal of 10\% of the data. Including inside-SAA sections increases the number of outliers somewhat, but this can be accounted for by using a non-normal noise model in the analysis (see Sect.~\ref{sec:error_models}). 

\begin{figure}
 \centering
 \includegraphics[width=\columnwidth]{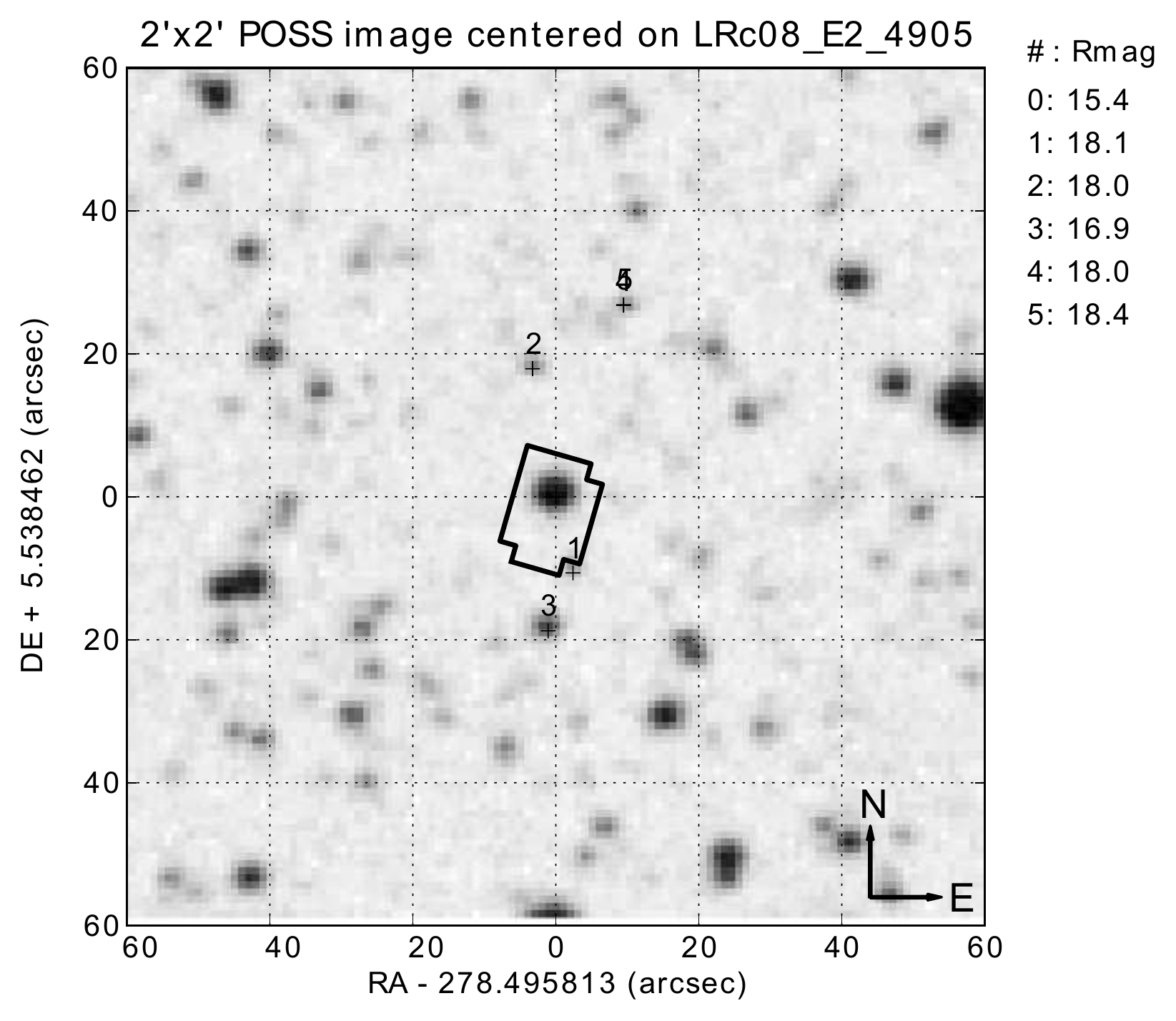}
 \caption{POSS image showing the surroundings of CoRoT-27 and the photometric aperture mask (solid line). One contaminating star (marked as 1) falls partially within the aperture mask, with an estimated contamination factor of $2.4\% \pm 0.95\%$.}
 \label{fig:phot_mask}
\end{figure}

We find one contaminating star within the CoRoT aperture mask (see Fig.~\ref{fig:phot_mask}) and estimate that it contributes $2.42\% \pm 0.95\%$ of the measured flux using a code developed by Bord\'e and Pasternacki \citep{Pasternacki2012}. 
We did not remove the contamination from the light curve before the combined light curve and RV analysis, but instead included contamination to the model with a normal prior based on the given estimate.

\begin{figure}
 \centering
%FIXME: Change me to PDF!
%  \includegraphics[width=\columnwidth]{./figs/full_lc.pdf}
 \includegraphics[width=\columnwidth]{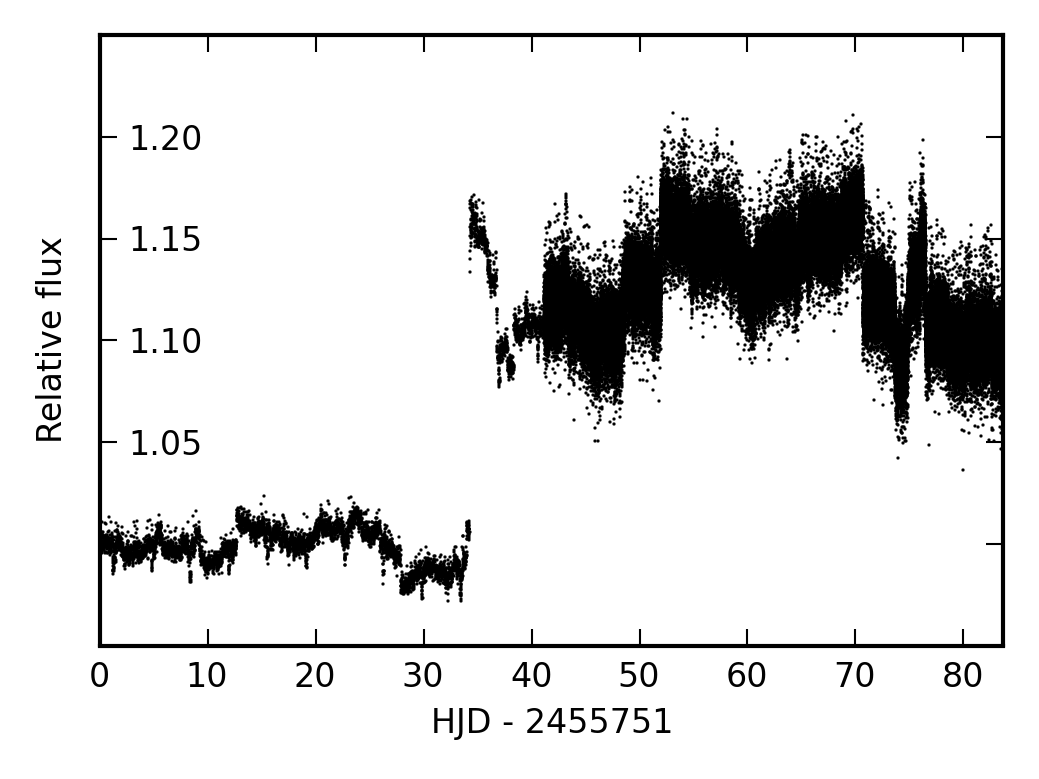}
 \caption{CoRoT-observed white light curve showing the long and short time cadence photometric data with extreme outliers (as marked by the CoRoT photometric pipeline) removed. The individual transits are visible by eye in the \lcad data.}
 \label{fig:full_lc}
\end{figure}

The final light curve used in the analysis is shown in Fig.~\ref{fig:full_lc}. The light curve consists of 6132~\lcad and 108471~\scad points (89~\% and 95~\% of all the data), features one large jump near the end of the \lcad data and several smaller jumps. The light curve contains 12~\lcad and 11~\scad transits. We did not attempt to carry out a global detrending to correct for the jumps and other systematics, but decided to use a local approach to estimate the systematic trends around each individual transit.

We estimate photometric point-to-point scatter of 4.3 and 12.5~ppt (parts per thousand) for \lcad and \scad, respectively, assuming independent and identically distributed (\iid) noise following a normal distribution, and 2.9 and 8.9~ppt assuming logistically distributed \iid noise (see the discussion about noise models in Sect.~\ref{sec:error_models}). The \lcad noise estimates are higher than expected for \iid noise, which can be explained by correlated (red) noise from instrumental and physical sources (stellar granulation, etc. See \citealt{Aigrain2009} for an overview of the CoRoT-specific error sources and \citealt{Pont2006} for an introduction to red noise in the context of photometric time series).

\subsection{Ground-based observations}
\subsubsection{Photometric follow-up}
Ground-based photometric follow-up of CoRoT candidates is verifying whether
a transit that is detected in CoRoT's large apertures is on a target
star instead of being caused by some nearby binary system \citep{Deeg2009}. An 18-min-long time series was acquired with the 1.2~m Euler
telescope during a transit on 29~September~2011 with excellent 0.5" seeing.
The seeing was poor, however, when the corresponding off-transit data
were taken a day later.  This strong difference in seeing meant that the
~1\% deep transit signal could not observed. However, none of the
nearby fainter stars showed any relevant brightness variation. The 
good on-transit images also did not show any very close contaminating stars
with a brightness that is sufficient (R$<$20.5) to cause the
transit signal. Contaminants causing a false alarm could therefore be
excluded for any distances greater than $\sim$1" from the target, giving a
very high probability that the transit does indeed arise on the target
star.

\subsubsection{Spectroscopic follow-up}
\begin{figure}
 \centering
 \includegraphics[width=\columnwidth]{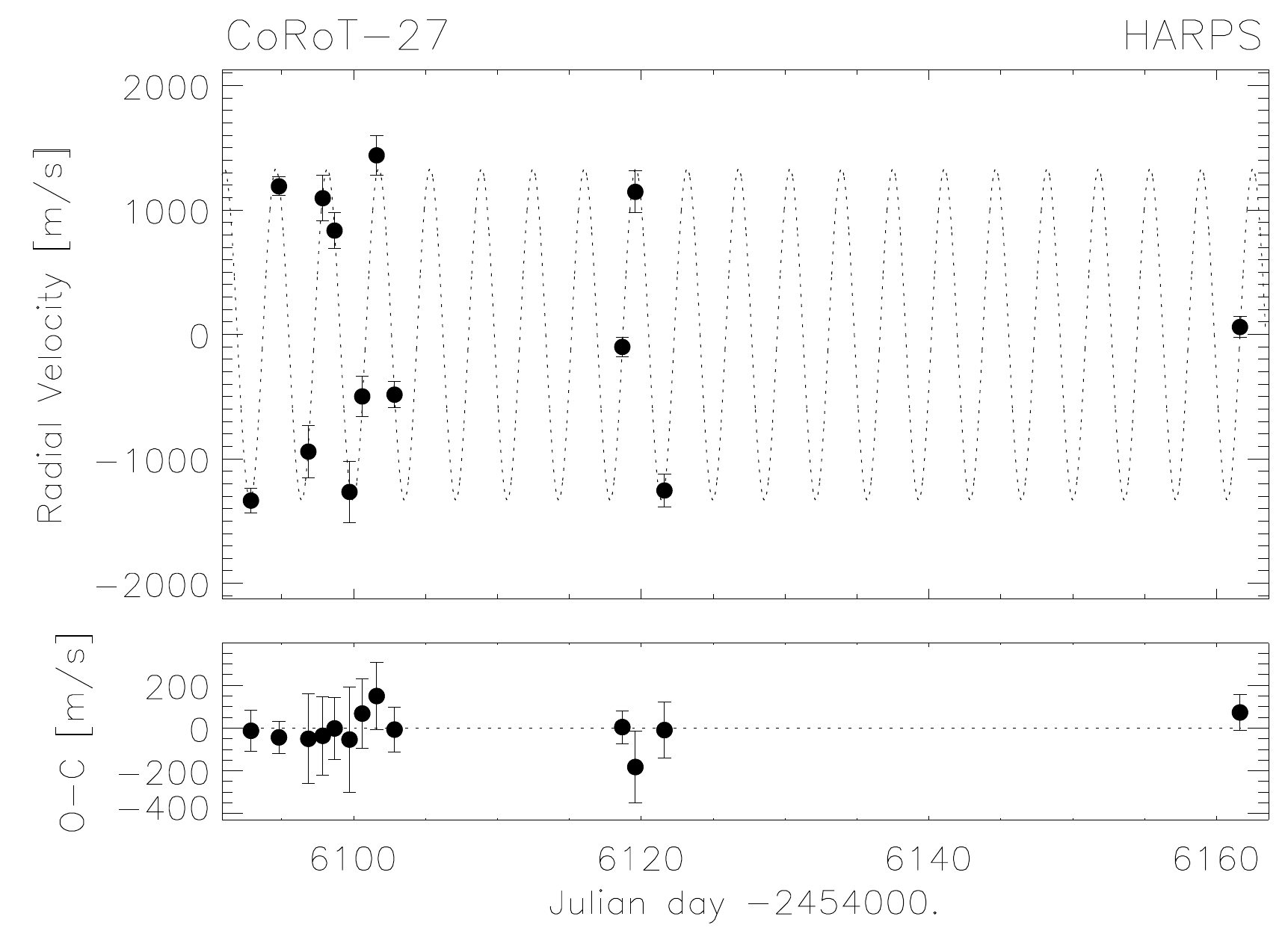}
 \caption{HARPS-observed radial velocities shown with a best-fit circular orbit (top) and residuals (bottom).}
 \label{fig:rv_time}
\end{figure}

High-precision radial-velocity observations were performed using HARPS at the 3.6~m telescope in La Silla Observatory, ESO, Chile (programme ID 188.C-0779). The thirteen measurements, shown in Fig.~\ref{fig:rv_time}, were carried out from 14 June to 21 August 2012, over 69 days. The HARPS mode with a spectral resolution of 110,000 was used. The signal-to-noise ratio of the observations varies from 1 to 5 at 550~nm during exposures of 1800~s, except the first two exposures of 3600~s.
The radial velocities (RV) were computed using cross-correlation with a G2 mask~\citep{Baranne1996, Pepe2002} after spectrum extraction with the HARPS pipeline. The cross-correlation function of CoRoT-27 shows a single peak with FWHM of 8.5~km/s.
From measured $B-V=0.96$ and HARPS calibration of the cross-correlation function, the estimated projected rotational velocity is $4.3\pm0.5$~km/s, in agreement with the spectroscopic analysis.
A simultaneous observation of the sky background allowed monitoring its evolution and impact on the stellar cross-correlation function, but all the observations performed on this faint star were unaffected by the sky background.

\begin{figure}
 \centering
 \includegraphics[width=\columnwidth]{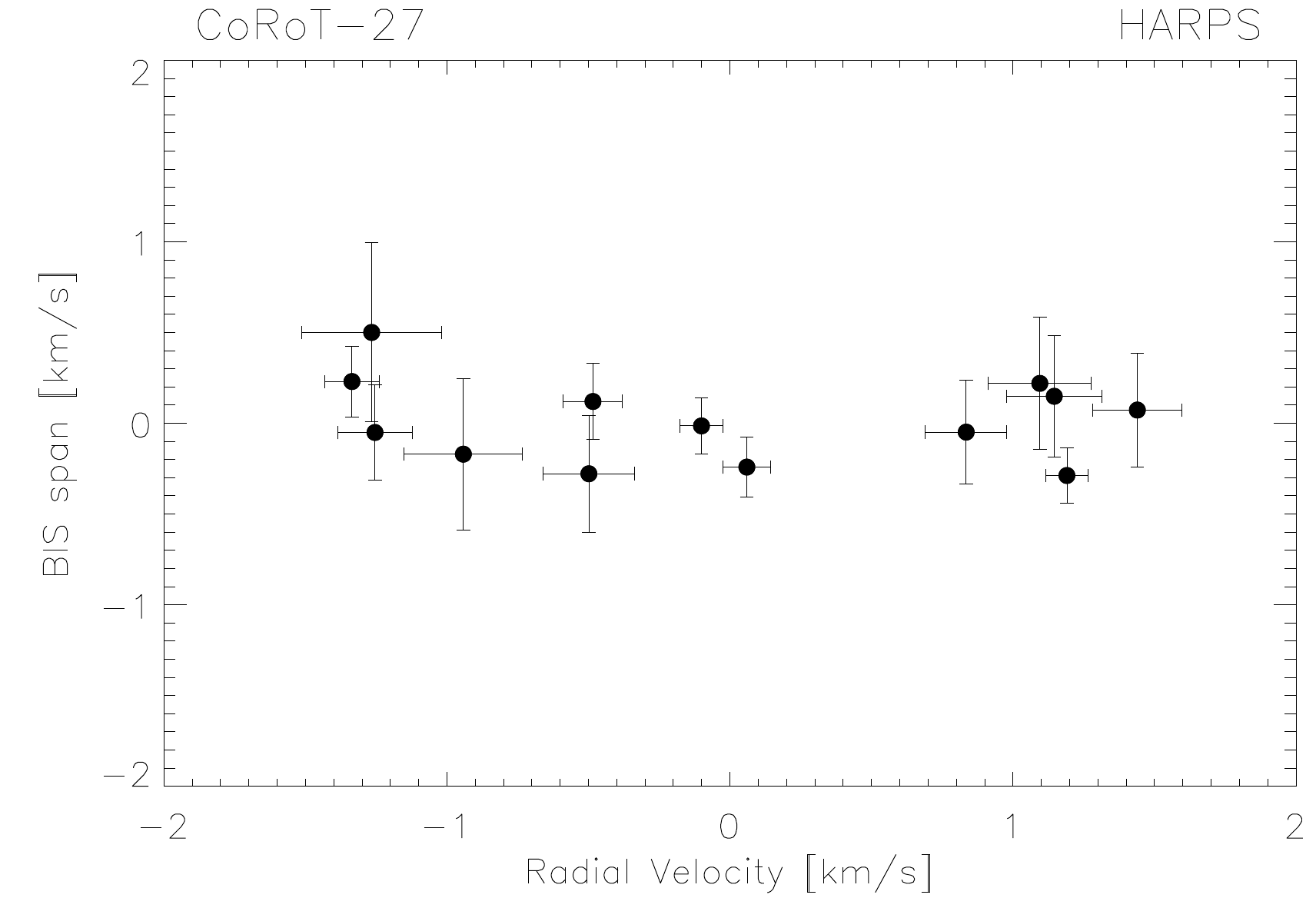}
 \caption{Bisector span as a function of radial velocity. The bisector spans show no correlation with the RV.}
 \label{fig:rv_bis}
\end{figure}

The HARPS data show a highly dispersed RV sequence, with a standard deviation of 1~km/s and peak-to-peak variation of 2.77~km/s.  The mean error on individual measurements is 140~m/s due to the faintness of the star.
The bisector spans, displayed in Fig.~\ref{fig:rv_bis}, show a standard deviation of 230~m/s without correlation with the RV. This is a good indication that the detected RV signal is due neither to photospheric activity nor to the blending effect of a background star.

\begin{table}[h] 
\caption{HARPS-observed RVs, their errors and bisector spans.}
\begin{tabular*}{\columnwidth}{@{\extracolsep{\fill}} lccr}
\toprule\toprule        
BJD - 2450000 & RV [km/s] & Error [km/s] & Bis [km/s] \\
\midrule
6092.85927 & -5.4157 & 0.0971 &  0.2299 \\
6094.80647 & -2.8896 & 0.0756 & -0.2877 \\
6096.85166 & -5.0221 & 0.2088 & -0.1702 \\
6097.84833 & -2.9852 & 0.1824 &  0.2197 \\
6098.67083 & -3.2454 & 0.1439 & -0.0488 \\
6099.71055 & -5.3461 & 0.2466 &  0.5005 \\
6100.59490 & -4.5786 & 0.1614 & -0.2782 \\
6101.59873 & -2.6407 & 0.1571 &  0.0727 \\
6102.84170 & -4.5642 & 0.1044 &  0.1200 \\
6118.67658 & -4.1807 & 0.0771 & -0.0146 \\
6119.58347 & -2.9342 & 0.1675 &  0.1487 \\
6121.60748 & -5.3344 & 0.1313 & -0.0508 \\
6161.61887 & -4.0204 & 0.0833 & -0.2415 \\
\bottomrule
\end{tabular*}
\label{tbl:rv}   
\end{table}

\section{Spectral analysis}
\label{sec:stellar_parameters}

As is now standard practice in the analysis of CoRoT stars, we used the co-added HARPS 
spectrum to derive the fundamental photospheric quantities of the planet hosting star
CoRoT-27, employing two different methods. The derived parameters include the effective 
temperature (\teff), surface gravity (\logg), metallicity ([M/H]), micro- and macro-turbulence 
velocities (\vmicro\ and \vmacro, respectively), and sky-projected stellar rotation 
velocity (\vsini).

A first method relies on the spectral analysis package \emph{Spectroscopy 
Made Easy} (SME, version 305), which calculates synthetic spectra of stars and fits them 
to observed high-resolution spectrum \citep{vp96}. It uses a non-linear least squares 
algorithm to solve the model atmosphere parameters (\teff, \logg, [M/H]), as well as the 
\vsini, \vmicro\ and \vmacro\ velocities. A set 
of LTE plane-parallel stellar model atmosphere grids \citep{kur93cd13,hau99,Gustafsson2008} 
are included with the SME distribution. 

We also used a customised IDL software suite to derive the fundamental photospheric 
parameters. We compared the co-added HARPS spectrum with a grid of theoretical 
model spectra from \citet{Castelli2004}, \citet{Coelho2005}, and \citet{Gustafsson2008},
using spectral features that are sensitive to the different photospheric parameters.

Briefly, we used the wings of the H$\alpha$ line to estimate the effective temperature of the 
star, and the Mg\,{\sc i} 5167, 5173, 5184~\AA lines, from the Ca\,{\sc i} 6162 and 6439~\AA lines, and from the 
Na\,{\sc i} D lines to determine its surface gravity. The metal abundance and v$_\mathrm{micro}$ 
were derived applying the method described in \citet{Blackwell1979}. We adopted the calibration 
from Bruntt et al. (2010) to estimate v$_\mathrm{micro}$. The projected rotational velocity v\,sin\,i 
was measured by fitting the profiles of several clean and unblended metal lines.

Consistent results were obtained with the two methods. The final adopted values for 
CoRoT-27 are \teff = 5900$\pm$120K, \logg = 4.4$\pm$0.10 dex, [M/H] = -0.1$\pm$0.1 dex, 
\vsini = 4.0$\pm$1.0 \kms, \vmicro = 1.3$\pm$0.5 \kms, and \vmacro = 1.9$\pm$0.5 \kms.

\section{Planet characterisation} 
\label{sec:analysis}

\subsection{Overview}
We use a Bayesian parameter estimation approach to characterise the planet and its orbit. The CoRoT-observed light curve and the HARPS-observed radial velocities are modelled jointly, and the information from stellar characterisation is used to set a prior on the stellar density. The posterior probability density estimates for the parameters are obtained using MCMC (Markov Chain Monte Carlo) sampling, and the final physical quantities are obtained by combining the parameter posteriors with the results form the stellar characterisation.
In addition to the basic characterisation of the planet, our analysis includes the search for a secondary eclipse. The search is carried out using a method based on Bayesian model selection between two competing models (with and without an eclipse), and is described in detail in \citet{Parviainen2012}.
The Bayesian approach facilitates the use of non-normal noise models in the analysis, and we investigate how the assumptions about the noise properties affect the outcomes of the parameter estimation and secondary eclipse search.

The analysis code is written in Python and Fortran and uses the common scientific Python libraries: NumPy, SciPy, Matplotlib \citep{Hunter2007}, and PyFITS. We model the transit shape using a version of the Gim\'enez transit model~\citep{Gimenez2006} optimised for efficient computation of large light curves\footnote{The code is freely available from {github.com/hpparvi/PyTransit}}.
The MCMC sampling of the posterior density is carried out using \textit{emcee} \citep{Foreman-Mackey2012}, a Python implementation of the affine invariant Markov chain sampler \citep{Goodman2010}. The sampler was chosen for its ability to efficiently sample correlated parameter spaces and for its self-adaptive nature, which  reduces the need to tweak the MCMC proposal distribution parameters by hand.

The light curve analysis is carried out for a subset of the data. We include a time span of 15.3~h centred on each individual transit into the analysis. Instead of detrending the light curve using a static polynomial fit to the out-of-transit data (as often done) or using the light curve filtered with running median or Savitzky-Golay filter, we model the background continuum using a separate Legendre series expansion for each transit. The Legendre polynomials are chosen for their orthogonality over the interval [-1,1], and the times around each transit are mapped to this interval. The coefficients of the Legendre series are free parameters in the MCMC analysis, which allows us to propagate the uncertainties in the background estimation to the physical parameter estimates.

This approach is feasible since we have multiple transit observations that allow us to reduce the degeneracy between the transit shape and background variations. The shape of the transit signal is constant over different transits, but the background variations are not. We carry out the analysis separately up to Legendre series expansion orders of 3, 4, and 5 to assess the effects from increasingly complex background modelling on the physical parameter estimates. 

We carry out the MCMC runs using 800 parallel Monte Carlo chains (walkers in emcee terms). We run the MCMC iteratively in batches of 400 steps, each batch iteration starting from the end state of the previous batch, until the parallel chains have converged to sample the true posterior distribution, and the median of each parameter is stable thorough an MCMC run. We use a thinning factor of 10 (a value chosen based on the average chain autocorrelation length), finally ending up with $800\times40 = 32000$ independent posterior samples. Thanks to our optimised transit-model code, the computation for a single MCMC batch iteration takes several tens of minutes on a single eight-core desktop computer, and the chains are found to converge after five to seven batch iterations.

The simulations are computed for three background models and four noise models in total, described below, ending up with a final set of 12 separate posterior estimates for each parameter.

\subsection{Bayesian parameter estimation}

The unnormalised posterior probability for a model parametrised by a parameter vector \pvec, given the light curve \Dlc and radial velocities \Drv, can be calculated as
\begin{equation}
 P(\pvec | \Dlc, \Drv) = P(\pvec) P(\Dlc | \pvec) P(\Drv | \pvec),
\end{equation}
where the first factor is the prior probability for \pvec, the second is the likelihood for the light curve data, given \pvec, and the last the likelihood for the radial velocity data, also given \pvec. The likelihoods are defined as
\begin{equation}
 P(\vec{D} | \pvec) = \prod_i P(e_i| \pvec) = \exp \left( \sum_i \ln P(e_i| \pvec) \right),
\end{equation}
where $e$ are the differences between the observed and modelled values (errors). The latter form is preferred for numerical stability, since the product over a large number of individual probabilities can easily lead to under- or overflows.

The exact form of $P(e|\pvec)$ depends on the assumptions made about the underlying noise distribution. We assume normally distributed \iid errors for the RV observations, but consider four noise distributions for the light curve data, described in more detail below.

\subsection{Parametrisation and priors}

The basic parametrisation of the combined light curve and RV model includes the 14 parameters listed in Table~\ref{table:models}. The mixture light curve noise models (discussed in Sect.~\ref{sec:error_models}) add four parameters, and the model used in the eclipse search adds one parameter. Finally, the coefficients of the Legendre series add from four to six parameters per transit, yielding from 96 to 144 additional parameters, in total, for 24 transits.

Since the affine invariant sampler is effective in sampling correlated parameter spaces, we parametrise the limb darkening using the two coefficients of the quadratic limb darkening law directly, instead of using their linear combinations. The quality of the light curve is not high enough to constrain the two degenerate limb darkening coefficients, and we use the theoretical models by \citet{Claret2011} to construct a normal prior N($\mu=0.27$, $\sigma=0.06$) on the quadratic (v) coefficient (see \citealt{Csizmadia2013} for a detailed overview of the effects of limb darkening on parameter estimates from transit light curves). The prior constrains the values of v, but is wide enough to account for the uncertainties in the stellar characterisation and theoretical stellar atmosphere modelling.

We let the eccentricity vary freely even when the RV data does not implicitly support significantly non-zero eccentricity. This is done in order to obtain robust estimates of the maximum eccentricity and the physical parameters derived from the analysis. Both the stellar density and semi-major axis estimates have been shown to be sensitive to the orbital eccentricity \citep{Kipping2010b}, and fixing the eccentricity to zero would lead to underestimated uncertainties for these parameters.

The flux contamination from a nearby star is also included in the model, and has a normal prior N$(\mu=2.42\%, \sigma=0.95\%)$ based on the estimate from the contamination analysis. Other parameters have uninformative uniform priors during the system characterisation and informative priors based on the posterior densities from the system characterisation during the eclipse search.

\begin{table}
\caption{Parametrisations used by the transit light curve (LC), radial velocity (RV), and eclipse (EC) models.}
\label{table:models}      
\centering  
% \begin{tabular}{\columnwidth}{@{\extracolsep{\fill}} lcccc}
\begin{tabular*}{\columnwidth}{@{\extracolsep{\fill}} lcccc}
\toprule\toprule
                    &Notation & LC & RV & EC \\
\midrule
Period	                 &\pper & X & X & X \\
Transit centre           &\ptc  & X & X & X \\
Impact parameter         &\pimp & X & X & X \\
Eccentricity             &\pec  & X & X & X \\
Argument of periastron   &\pom  & X & X & X \\
\\
Limb darkening coefficients &u,v& X &   &   \\
Planet-star radius ratio &\prr  & X &   & X \\
Reciprocal of half T$_1$ duration\tablefootmark{a}
                         &\prtd & X &   & X \\
Contamination            &\pcf  & X &   & X \\
Long cadence noise std   &\pelc & X &   & X \\
Long cadence noise tail std\tablefootmark{b} 
                         &\pelt & X &   &   \\
Long cadence noise mix ratio\tablefootmark{b}
                         &\pelr & X &   &   \\
Short cadence noise std  &\pesc & X &   & X \\
Short cadence noise tail std\tablefootmark{b}
                         &\pest & X &   &   \\
Short cadence noise mix ratio\tablefootmark{b}
                         &\pesr & X &   &   \\
\\
RV systemic velocity     &\prvz &   & X &   \\
RV semi-amplitude        &\prva &   & X &   \\
\\
Planet-star flux ratio   &\pfr  &   &   & X \\
\bottomrule
\end{tabular*}
\tablefoot{
\tablefoottext{a}{See \citet{Kipping2010b}.}
\tablefoottext{b}{Included in the mixture noise models.}
}
\end{table}

\subsection{Noise models}
\label{sec:error_models}
We consider four different zero-centred noise models, all without correlated noise. First, we have the normal distribution
\begin{equation}
 P_\mathrm{n}(e|\sigma) = \frac{1}{\sigma \sqrt{2\pi}} \exp \left (-\frac{e^2}{2\sigma^2} \right ),
\end{equation}
and the more heavily tailed logistic distribution
\begin{equation}
 P_\mathrm{l}(e|s) = \frac{\exp (- \frac{e}{s})}{s(1 + \exp (-\frac{e}{s})}.
\end{equation}
The normal distribution works as a standard against which we compare the other models. The logistic distribution was chosen as the second two-parameter model since its shape is close to the normal distribution, but it includes heavier tails that make the analysis less sensitive to outliers. Student's distribution would be the next step from logistic distribution, allowing for enhanced flexibility in the distribution shape (and introducing one additional parameter). However, logistic distribution is significantly faster to evaluate, and Student's distribution was not included in the analysis.

Next, we consider two mixture models. The mixture models are linear combinations of two distributions, where the first distribution models the body of the distribution and the other one the tails. The mixture normal-normal distribution models the noise with two normal distributions with a common mean but different $\sigma$
\begin{equation}
 P_\mathrm{n,n}(e|\sigma_1, \sigma_2, \beta) =  (1-\beta) P_\mathrm{n}(\sigma_1) + \beta P_\mathrm{n}(\sigma_2),
\end{equation}
where $\beta$ is the mixing factor weighting the two distributions.
The mixture normal-Cauchy distribution models the noise with a mixture of normal and Cauchy distributions as
\begin{equation}
 P_\mathrm{n,C}(e|\sigma, s, \beta) = (1-\beta) P_\mathrm{n}(\sigma) + \beta \left( \pi s \left (1+ \frac{e^2}{s^2} \right) \right)^{-1},
\end{equation}
where $s$ is the Cauchy distribution's half width at half maximum.
The normal distribution is again used to model the main body of the distribution, while the Cauchy distribution adds long tails.

\begin{figure}[t]
 \centering
 \includegraphics[width=\columnwidth]{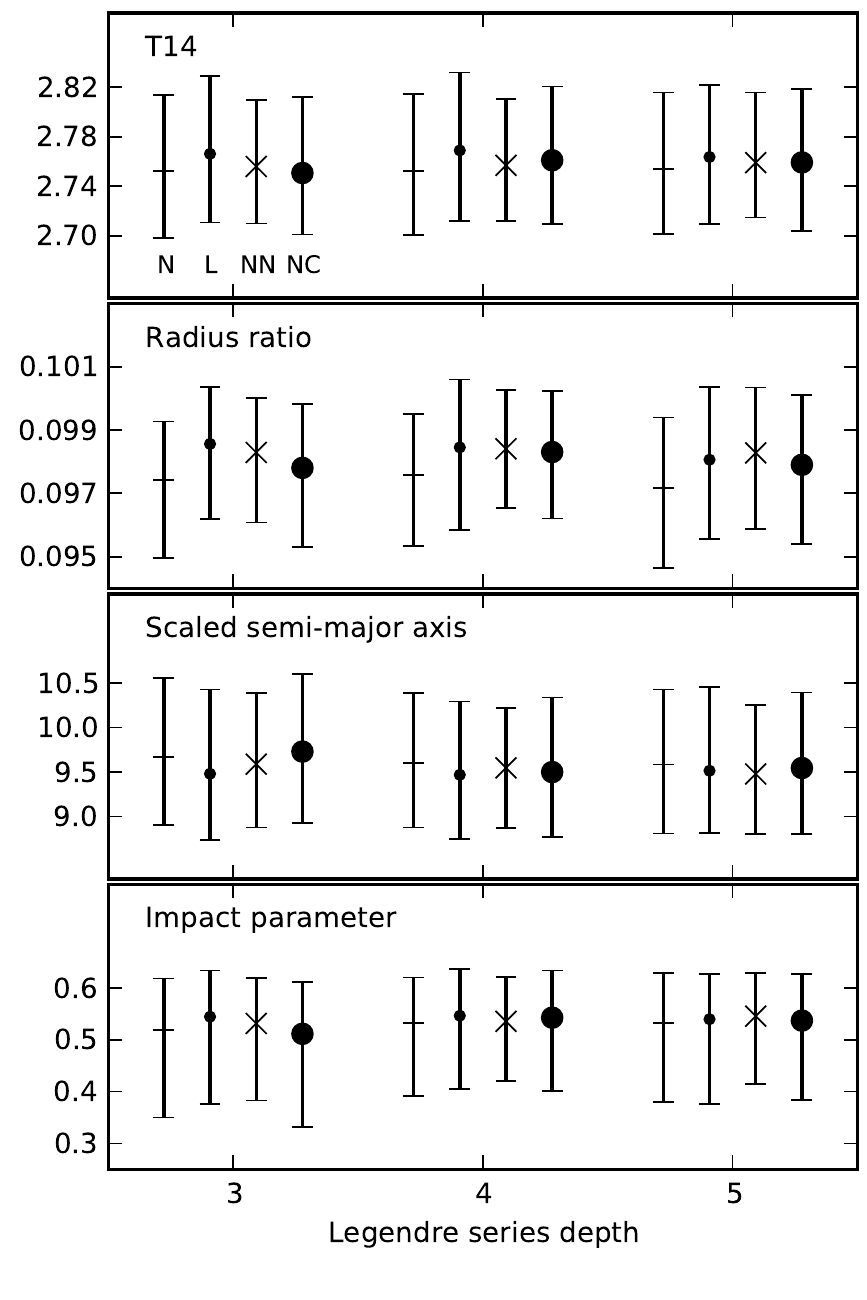}
 \caption{Parameter estimates for different noise and background models. The Legendre series depth tells the maximum order of Legendre polynomials used in the continuum fitting. N corresponds to normal error model, L to logistic, NN to combined normal-normal model, and NC to combined normal-Cauchy model.}
 \label{fig:parameter_estimates}
\end{figure}

\begin{figure}
 \centering
 \includegraphics[width=\columnwidth]{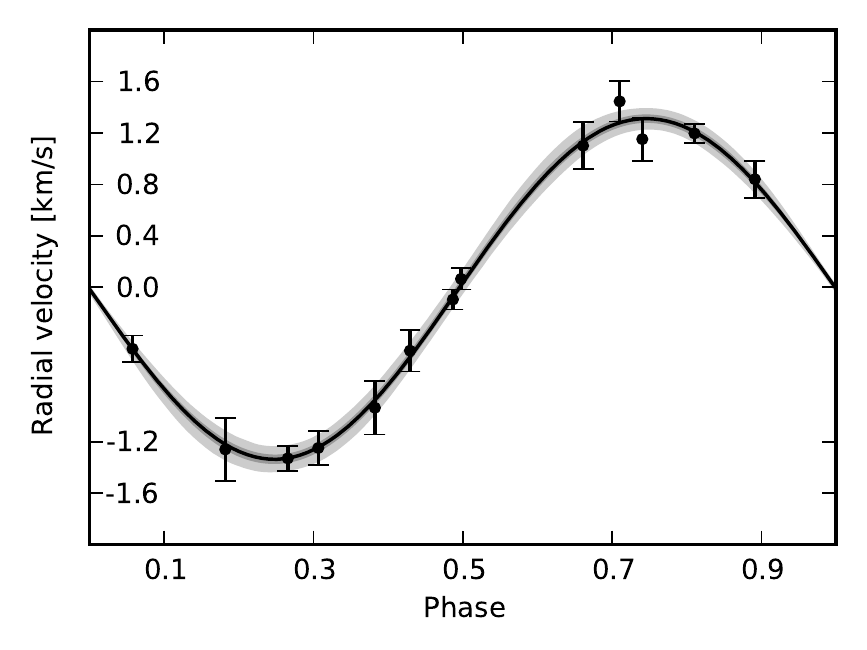}
 \caption{Phase-folded radial velocity points with the median (black line) and the 95\% limits (shaded area) of the predictive distribution.}
 \label{fig:rv_fit}
\end{figure}

\begin{figure}
 \centering
 \includegraphics[width=\columnwidth]{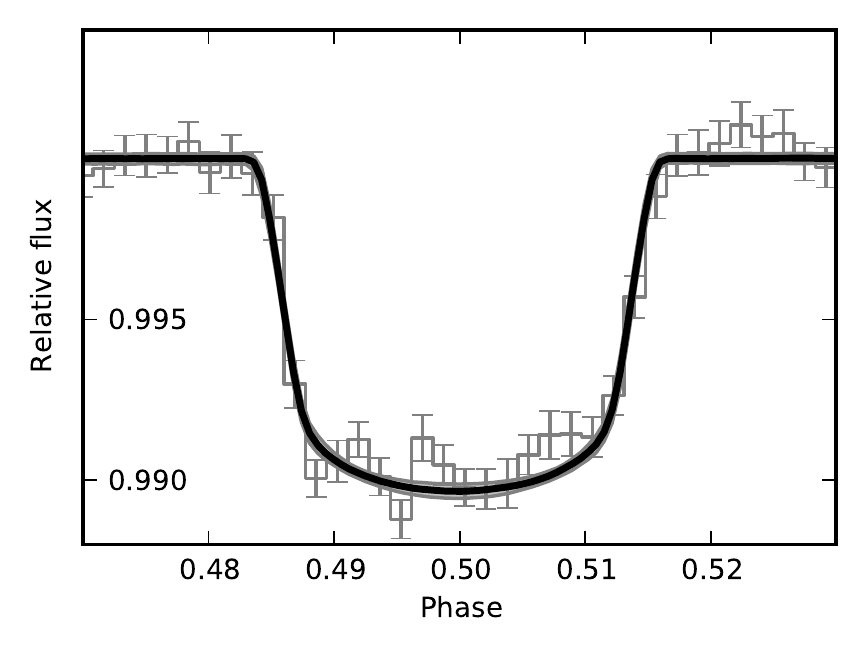}
 \caption{Phase-folded and binned light curve with the median (black line) and the 95\% limits (shaded area) of the predictive distribution.}
 \label{fig:lc_fit}
\end{figure}

\begin{figure}
 \centering
 \includegraphics[width=\columnwidth]{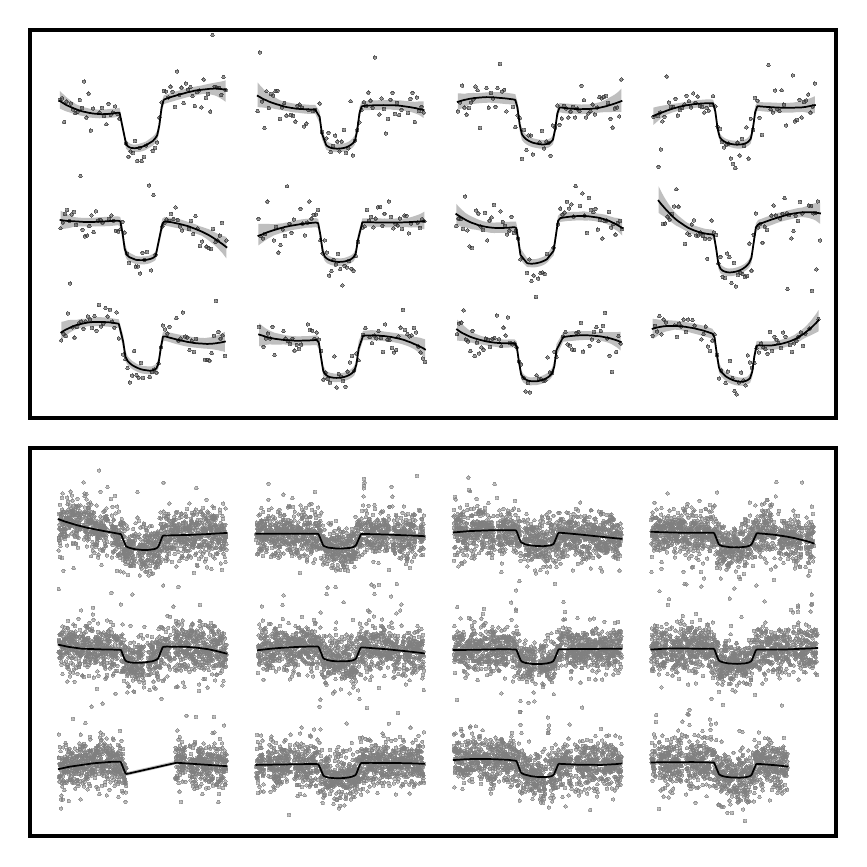}
 \caption{Individual transits with the median (black line) and the 95\% limits (shaded area) of the predictive distribution. Upper figure shows the \lcad data and lower the \scad data.}
 \label{fig:ind_lcs}
\end{figure}

\begin{figure}
 \centering
 \includegraphics[width=\columnwidth]{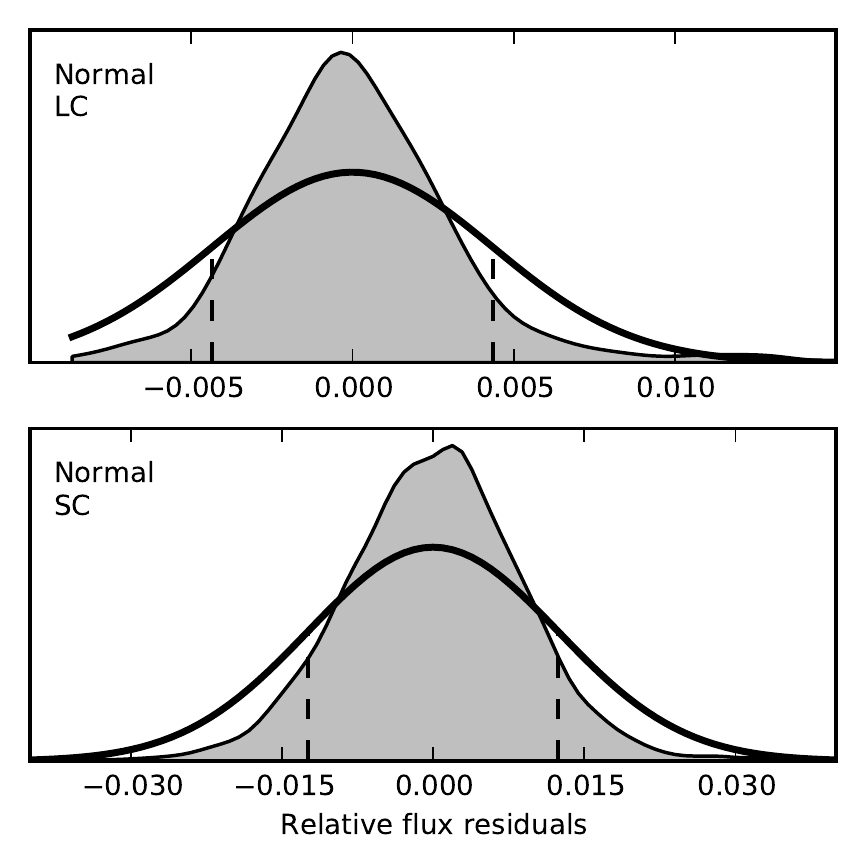}
 \includegraphics[width=\columnwidth]{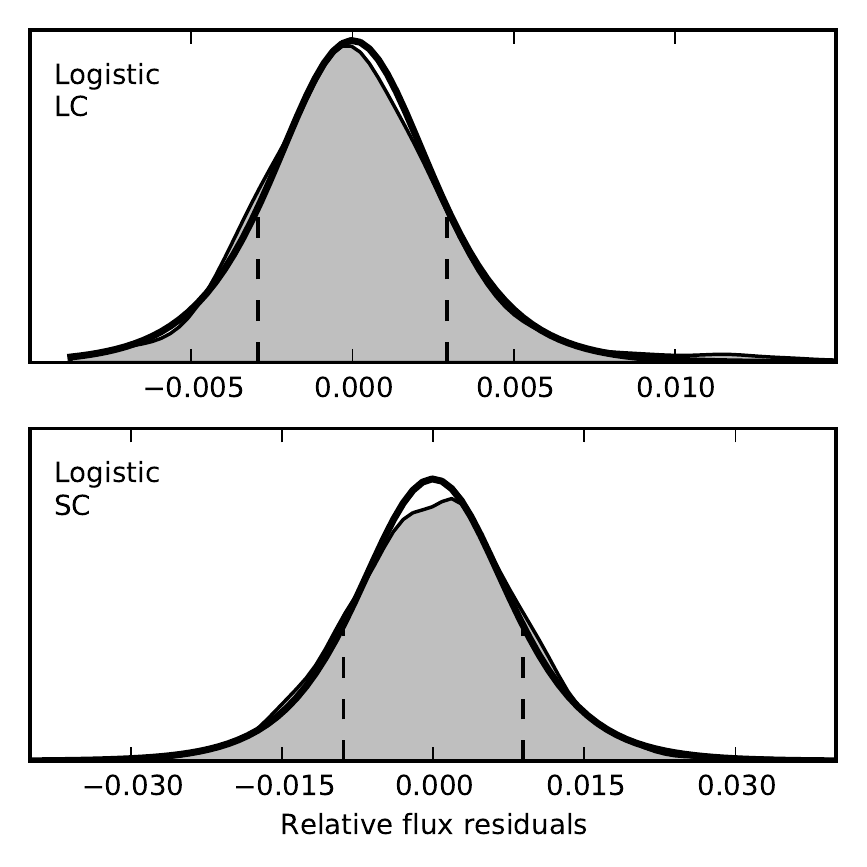}
 \caption{Residual distributions for the normal and logistic noise models. The shaded area shows the residual distribution and the thick line the fitted noise model. The long cadence data features a long positive tail missing from the short cadence data.}
 \label{fig:residuals}
\end{figure}

\begin{figure}
 \centering
 \includegraphics[width=\columnwidth]{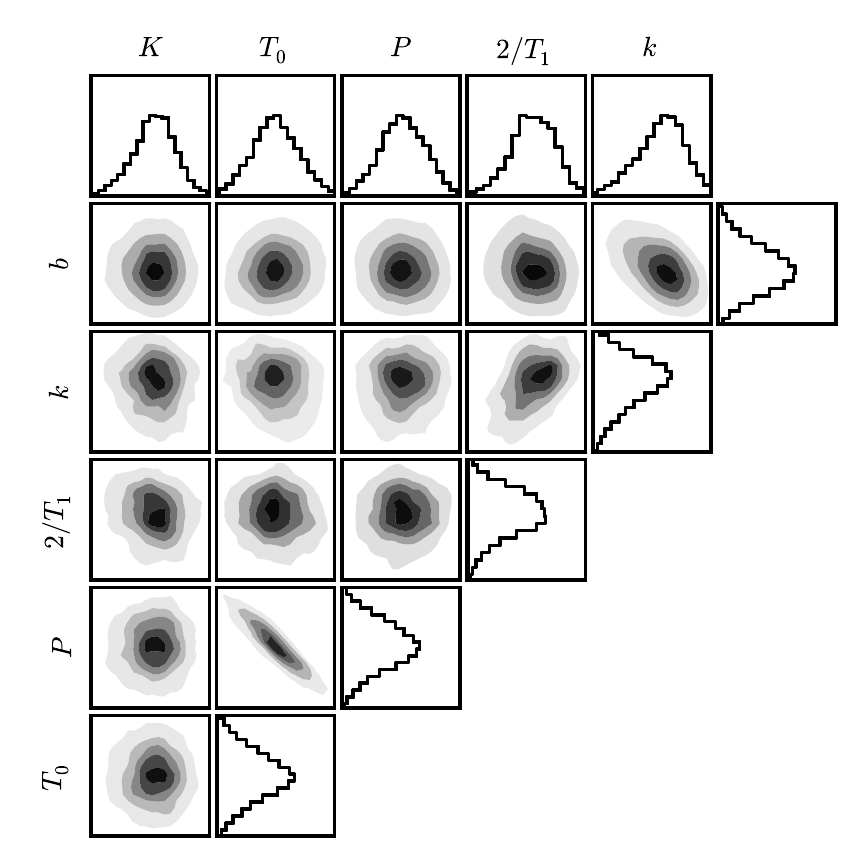}
 \caption{Marginal distributions for the fitting parameter set and the parameter-parameter correlations. The estimates are found in Table \ref{table:parameters}.}
 \label{fig:parameter_distributions}
\end{figure}

\subsection{Parameter estimation results}
We show the parameter estimates from the MCMC runs that employ all combinations of noise and background models for transit duration, radius ratio, scaled semi-major axis, and impact parameter in Fig.~\ref{fig:parameter_estimates}. The runs are labelled as LDE, where D is depth of the Legendre expansion ($\mathrm{D} \in \{3,4,5\}$), and E the noise model ($\mathrm{E} \in \{N,L,NN,NC\}$). The estimates show very little sensitivity to the chosen noise or background model, and we adopt the values from the simplest background model and the simplest non-normal noise model; that is, from the L3L MCMC run. The final parameter estimates, corresponding to the posterior median values and the 68\% confidence limits, are listed in Table~\ref{table:parameters}, and the posterior distributions and correlations are shown in Fig.~\ref{fig:parameter_distributions}. We omit the limb darkening coefficients from Fig.~\ref{fig:parameter_distributions}, since the normal prior on the v coefficient reduces the correlations to 
inconsequential.

We show the median and the 95\% limits of the predictive distribution for the RV model in Fig.~\ref{fig:rv_fit} and for the light curve model in Fig.~\ref{fig:lc_fit}. The radial velocities in Fig.~\ref{fig:rv_fit} have the systemic velocity removed, and the light curve data shown in Fig.~\ref{fig:lc_fit} are normalised using the best-fit background model, phase folded, and binned for visualisation purposes. Finally, we show the individual transits from the L3L run (and the 95\% predictive distribution limits) in Fig.~\ref{fig:ind_lcs}. 

Based on Fig.~\ref{fig:parameter_estimates}, the noise model only plays a minor role in the parameter estimation. However, this may not be the case in a model comparison problem, such as the search for secondary eclipses described in Sect.~\ref{sec:secondary_search}. We show the light curve residual distributions and the estimated noise distributions for the normal and logistic noise models in Fig.~\ref{fig:residuals}. The logistic model reproduces the true error distribution with much higher accuracy than the normal model. The two mixture models yield close-to equal fits from the logistic model, but at the price of adding four parameters.

\begin{table}
\centering
\small
\caption{Planet and star parameters.}            
\begin{minipage}[t]{\columnwidth} 
\setlength{\tabcolsep}{1.0mm}
\renewcommand{\footnoterule}{}   
\begin{tabular*}{\columnwidth}{@{\extracolsep{\fill}} llll}        
\toprule\toprule
\multicolumn{4}{l}{\emph{Ephemeris}} \\
\midrule
Planet orbital period & $P$ & [days] &  $3.57532 \pm 6 \times 10^{-5}$ \\
Transit epoch & $T_0$ & [HJD] & $2455748.684 \pm 0.001$\\
Transit duration & $T_{14}$ & [h] & $2.77 \pm 0.06$ \\
\\
\multicolumn{4}{l}{\emph{Results from radial velocity observations}} \\
\midrule   
Orbital eccentricity &$e$ & &  $< 0.065$ (99\% conf.) \\
RV semi-amplitude &$K$ &[\ms] & $ 1326 \pm 33$ \\
Systemic velocity  &$V_{r}$ &[\kms] & $-4.086 \pm 0.024$ \\
\\
\multicolumn{4}{l}{\emph{Fitted transit parameters}} \\
\midrule
Radius ratio &$k$ && $ 0.099 \pm 0.002$ \\
Limb darkening coeff. &$u$ && $0.25 \pm 0.12$ \\
Impact parameter &$b$ && $0.54^{+0.09}_{-0.17}$\\
\\
\multicolumn{4}{l}{\emph{Deduced transit parameters}} \\
\midrule
Scaled semi-major axis &$a/R_{*}$ && $9.48^{+0.95}_{-0.75} $ \\
$M^{1/3}_{*}/R_{*}$ &&[solar]& $0.97^{+0.10}_{-0.08}$ \\
Stellar density &$\rho_{*}$ &[$g\;cm^{-3}$] & $1.24^{+0.42}_{-0.26}$\\
Inclination &$i$ &[deg] & $86.7^{+1.2}_{-0.87}$ \\
\\
\multicolumn{4}{l}{\emph{Spectroscopic parameters }} \\
\midrule
Effective temperature &$T_{\mathrm{eff}}$ &[K] & $5900 \pm 120$ \\
Surface gravity &log\,$g$ &[dex]&  $4.4 \pm 0.1$  \\ % logg deduced : logg = 4.42
Metallicity &$[\rm{Fe/H}]$ &[dex]&  $0.1 \pm$ 0.1\\
{\vsini} &&[\kms]& $4.0 \pm$ 1.0\\
Spectral type & & &G2 \\
\\
\multicolumn{4}{l}{\emph{Stellar and planetary physical parameters}} \\
\midrule
Star mass & \smass &[\Msun]&  $1.05 \pm 0.11$ \\ 
Star radius & \srad &[\Rsun]&  $1.08^{+0.18}_{-0.06}$  \\ 
Age of the star &$t$ &[Gyr] & $ 4.21 \pm 2.72$ \\
Semi-major axis& $a$ &[AU] & $ 0.0476 \pm 0.0066$ \\
Planet mass &$M_{\mathrm{p}}$ &[M$_J$ ]$^d$ &   $10.39 \pm 0.55$ \\
Planet radius &$R_{\mathrm{p}}$ &[R$_J$]$^d$  &  $1.007 \pm 0.044$ \\
Planet density &$\rho_{\mathrm{p}}$ &[$g\;cm^{-3}$] &  $12.60_{-1.67}^{+1.92}$ \\
Eq. temperature &$T_{\mathrm{eq}}$ &[K] & $1500 \pm 130$ \\
\bottomrule       
\vspace{-0.2cm}
%add if necessary
% \footnotetext[1]{$I(\mu)/I(1)=1-u(1-\mu)$, where $I(1)$ is the specific intensity at the centre of the disk and $\mu=\cos{\gamma}$,$\gamma$ being the angle between the surface normal and the line of sight;}
%% next footnote IF orbit is eccentric
% \footnotetext[2]{$a/R_{*}=\frac{1+e \cdot \cos{\nu_{1}}}{1-e^{2}} \cdot \frac{1+k}{\sqrt{1-\cos^{2}({\nu_{1}+\omega-\frac{\pi}{2}})\cdot \sin^{2}{i}}}$, where $\nu_{1}$ is the true anomaly measured from the periastron passage at the transit egress (see \citealt{Gimenez2009}) .}
%\footnotetext[3]{$b=\frac{a \cdot \cos{i}}{R_{*}$}
% \footnotetext[3]{$b=\frac{a \cdot \cos{i}}{R_{*}} \cdot \frac{1-e^{2}}{1+e \cdot \sin{\omega}}$}
\footnotetext[4]{Radius and mass of Jupiter taken as 71492 km and 1.8986$\times$10$^{30}$ g.}
%\footnotetext[5]{zero albedo equilibrium temperature for an isotropic planetary emission.} 
\end{tabular*}
\end{minipage}
\label{table:parameters}  
\end{table}

\section{Discussion}
\label{sec:discussion}

\subsection{Secondary eclipse search}
\label{sec:secondary_search}
\subsubsection{Overview}
We search for secondary eclipses using the method based on Bayesian model selection described in \citet{Parviainen2012}. In summary, we integrate the posterior density over the whole parameter space for the models with ($M_1$) and without ($M_0$) eclipse signals. The models use the same parametrisation as the combined transit and RV analysis, but the eclipse model also includes a planet-star flux ratio (defined as the ratio between the fluxes per projected surface area element, so that eclipse depth $\Delta F = fk^2$) as a new parameter. The parameter priors are derived from the posterior densities of the L3 normal and logistic MCMC runs, and for the flux ratio we assign a Jeffreys' prior \citep{Jeffreys1946} from $10^{-3}$ to $5 \times 10^{-2}$.

Since the planet characterisation MCMC runs have shown that the heavy-tailed noise distributions model the point-to-point scatter better than the normal distribution, we improve upon the approach by \citet{Parviainen2012} by calculating the model posteriors also for the logistic noise distribution. We do not consider the two mixture models since the improvement in the modelling of the distribution was not substantial enough to justify the introduction of four additional parameters.

\subsubsection{Results}
We find very low Bayes factors, $B_{10}$, of 0.0069 and 0.013 in favour of the eclipse model for the normal and logistic noise models, respectively. We also carried out Bayes factor mapping as described in \citet{Parviainen2012}, and find maximum Bayes factors below two for both noise models. Thus, we can rule out a detectable secondary eclipse in the light curve data.

\subsubsection{Sensitivity tests}
We tested the sensitivity of the eclipse search method by injecting eclipse signals of various depths to the data. We find that the cut-off between a detectable and undetectable signal is sharp for both noise models, but that the use of the logistic noise model increases our sensitivity to the signal. 
Our tests show detection thresholds for flux ratio of 10\% and 9\% for the normal and logistic models, respectively. This is significantly higher than the flux ratios that can be expected for the system. While the threshold is on a similar scale for both noise models, the models show a difference in the detection sensitivity when going above the threshold. For $f=0.12$, for example, we obtain a maximum $B_{10}$ of 5.4 for the normal and 44.7 for the logistic model, respectively.

The differences in sensitivity between the two models can be explained by Fig. \ref{fig:residuals}. While the difference between the models is minor in the parameter estimation, the maximum likelihood obtained for the logistic model is $10^{2080}$ higher than the one for the normal model. The normal model assumes larger scatter to the data than the logistic, and yields a lower significance for small signals close to the noise limit.

\subsection{Search for additional planets}
We carried out a search for additional planets from the photometry and radial velocities after removing the best-fitting transit and RV models of CoRoT-27b from the data. No new significant planet signals were detected.

\subsection{Structure and composition of CoRoT-27b}
\label{sec:structure_and_composition}

CoRoT-27b is a massive hot Jupiter with a mass of 10.39~\mjup, a radius of 1.007~\rjup, and an inferred density of 12.60~\gcm.
Only a few other giant planets share the same parameter space: HAT-P-20b \citep{Bakos2011b}, CoRoT-20b \citep{Deleuil2012a}, WASP-18b \citep{Hellier2009},
XO-3b \citep{JohnsKrull2008}, and the most similar Kepler-75b \citep[KOI-889b, ][]{Hebrard2013} with a density of 11~\gcm for a mass of 9.9~\mjup
and a radius of 1.03~\rjup. Among these, HAT-P-20b and CoRoT-20b have been modelled and are believed to contain large amounts of heavy elements in
their interior.

\begin{figure}
 \centering
 \includegraphics[width=\columnwidth]{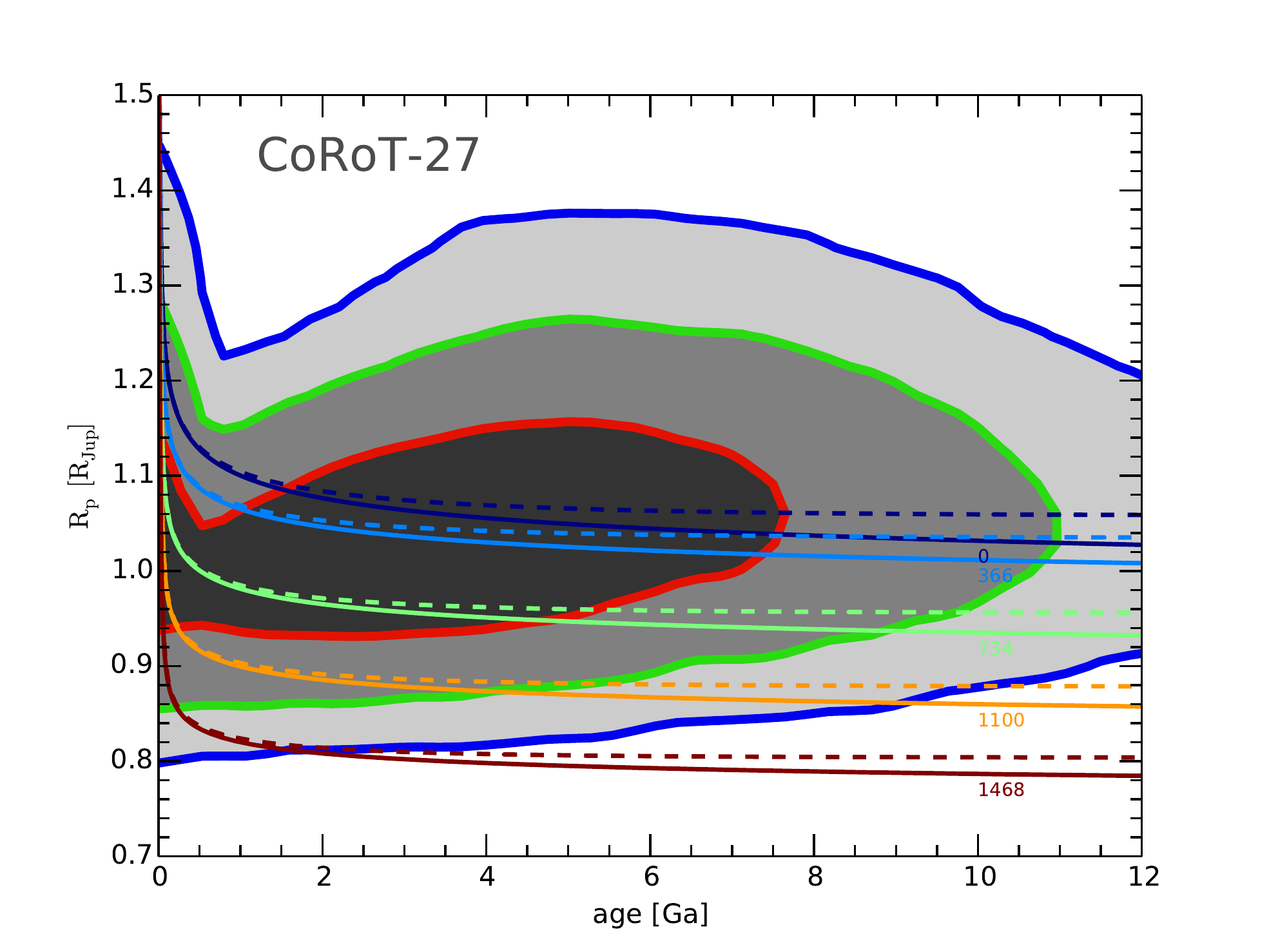}
 \caption{Transit radius of CoRoT-27b as a function of age, as computed by SET. The 68.3\%, 95.5\%, and 99.7\% confidence regions are denoted by black, 
  dark grey, and light grey areas, respectively. The curves represent the thermal evolution of a 10.39~\mjup planet with an equilibrium temperature 
  of 1500~K. Text labels indicate the amount of heavy elements in the planet (its core mass, in Earth masses). Dashed lines represent planetary evolution models
  for which 0.25\% of the incoming stellar flux is dissipated into the core of the planet, whereas plain lines do not account for this dissipation (standard models).}
 \label{fig:c27_struct}
\end{figure}

Combined stellar \citep[PARSEC,][]{Bressan2012} and planetary \citep[CEPAM,][]{Guillot1995, Guillot2010a} evolution models of the CoRoT-27b system were
calculated with SET \citep{Guillot2011, Havel2011}. Posterior probabilities of the planet's bulk composition were computed with an MCMC algorithm using a likelihood based on stellar (resp. planetary) observables [Fe/H], $T_{eff}$, $\rho_\star$, $\log g_\star$ (resp. transit depth $\Delta f$ and radial velocity $K$), and grids of models for the star and the planet. The results are presented in terms of planetary radii as a function of age in Fig.~\ref{fig:c27_struct}. Solutions within 68.3\%, 95.5\%, and 99.7\% confidence regions are shown with different colours.
%NOTE: Different colours don't help in an greyscale plot

Planetary evolution models are calculated in two cases: the ``standard'' case (which assumes that the thermal evolution of the planet is only a consequence
of the loss of its primordial entropy through the irradiated planetary atmosphere, Fig.~\ref{fig:c27_struct}), and one in which a fraction of the incoming
stellar flux is converted into kinetic energy and then dissipated at the centre of the planet \citep[see][for a
discussion]{Guillot2002, Guillot2006}. In addition, for each of these cases, two classes of models are considered: one in which the planet is made
of a central rocky core and a solar-composition envelope; and at the other extreme (not shown), the second class considers that the heavy elements are only present
in the envelope (using an equivalent helium mass fraction $\rm Y_{equiv}$ for the SCVH EOS). We clearly do not know which one of these assumptions is
closer to reality, but they represent the two extremes for the planetary radii as a function of the heavy element content
(see \citealt{Guillot2006,Baraffe2008, Ikoma2006}).

Both the ``standard'' and ``dissipated-energy'' models provide solutions for the planetary radius that match the available constraints. In fact, for a given
age, the difference in planetary radius between the two models is small compared to the uncertainty reported in Table 4 (about half, or 0.02~\rjup, for the $1\sigma$ confidence region shown in Fig.~\ref{fig:c27_struct}). We therefore consider global solutions that mix both cases\footnote{Numbers reported for
the bulk composition of the planet come from MCMC 1-D distributions, so may seem a bit inconsistent.}. For the first class of
models, we infer a core mass of $366_{-241}^{+267}\rm\,M_{\oplus}$, which translates into a heavy element mass fraction of $0.11_{-0.07}^{+0.08}$.
For the second class of models, we infer a heavy element mass fraction of $0.07_{-0.05}^{+0.06}$, which translates into a heavy elements mass of
$219_{-149}^{+206}\rm\,M_{\oplus}$. As expected \citep{Baraffe2008}, mixing the heavy elements in the envelope significantly reduces the amount needed to match
the observed radius: about a 40\% reduction, or $\rm 147\,M_{\oplus}$.
Interestingly, models with no heavy elements at all cannot be excluded, and they match the $1\sigma$ confidence region on Fig.~\ref{fig:c27_struct} well. Also, putting all heavy
elements in a massive rocky core allows for a much higher $1\sigma$ limit of about 2 $\rm M_{Jup}$ (against 1.35~\mjup for the $\rm Y_{equiv}$
class of models).
While qualitatively in line with what has been found for irradiated transiting giant planets \citep{Miller2011}, CoRoT-27b may require a surprisingly high amount
of heavy elements. In fact, the core class of models has demanded using the \citet{Zapolsky1969} zero-temperature EOS to properly span the possible
core masses matching the constraints. This EOS should be accurate enough within 1-20\% range for rocky materials \citep[e.g.][]{Fortney2007, Mordasini2012a}. Formation of such a planet with a very high amount of heavy elements remain uncertain \citep{Mordasini2012a, Mordasini2013} and would favour solutions with higher planetary radii in the case of CoRoT-27b.

\subsection{Blending}
\label{sec:blending}

\begin{figure}
 \centering
 \includegraphics[width=\columnwidth]{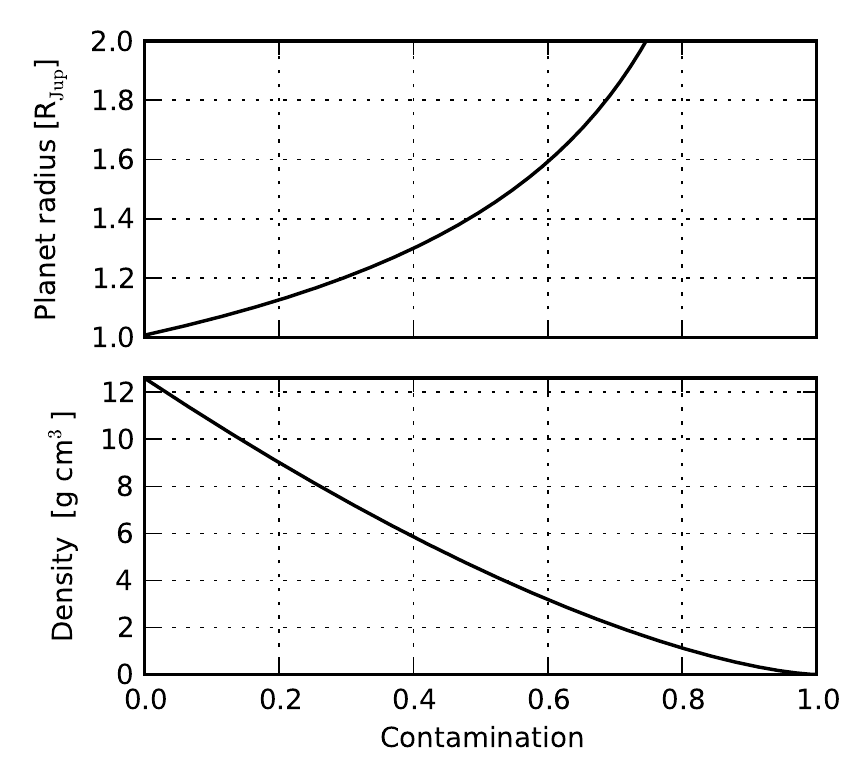}
 \caption{Effects of possible blending on the planet radius and density.}
 \label{fig:blending}
\end{figure}

While the photometric follow-up excludes contaminants located~$>1$\arcsec{} from the main target, we cannot rule out the possibility of a closer contaminant, which would have the effect of diluting the transit and thus lead to an under-estimated planet radius \citep[as with Kepler-14b,][]{Buchhave2011a}. The true planet-to-star area ratio $A$ depends on the contamination factor $c$ (the fraction of the flux in the aperture contributed by other stars than the one being occulted) and the observed (blended) planet-to-star area ratio $A_b$ as: $A = A_b(1-c)^{-1}$. Consequently, the planet-to-star radius ratio $k$ (hence the planet radius) scales as $R_{\rm p} \propto k \propto A^{0.5} \propto (1-c)^{-0.5}$, and the planet density as $\rho \propto A^{-1.5} \propto (1-c)^{1.5}$. Figure~\ref{fig:blending} illustrates the dependence of $R_{\rm p}$ and $\rho$ on contamination factors ranging from 0 to 0.99.

A coarse estimate for the maximum contamination can be obtained based on the properties of the current massive planet population. The radii of the known transiting exoplanets with $\pmass > 5~\mjup$  vary from 0.86~to~1.28~\rjup. Thus, assuming that the true CoRoT-27b radius lies roughly within this range, we would obtain a maximum contamination factor of $\sim$0.4, which would correspond to a minimum density of $\sim$6~\gcm.  

% Further constraints could in principle be set based on the stellar density. Since the transit duration is a direct measurable, increasing the radius ratio affects the orbital semi-major axis and %stellar density estimates.

\subsection{Age of the system}
Theoretical modelling yields two age estimates for the star: a young star of $3.2_{-0.8}^{+0.6}$~Myr, or an older star of $4.2\pm2.7$~Gyr. The young solution can 
be ruled out based on planetary contraction models, while the older solution can explain the measured mass density and radius for a range of planetary core masses.

\subsection{Stellar rotation period}
An attempt was made to measure the stellar rotation period using the method by \citet{MCQuillan2013} based on the autocorrelation function (ACF). The analysis was carried out using a jump-corrected version of the light curve, but no periods could be detected reliably. The result is not surprising given the amount of systematics in the light curve (see Fig.~\ref{fig:full_lc}).

\subsection{Tidal evolution}
\label{sec:tidal_evolution}

The system of CoRoT-27---where a massive planet is moving close to the star and for which high-quality parameters can be determined from both the transits and the spectroscopic observations---is particularly well suited to the study of tidal evolution. The starting point in any tidal evolution study is the choice of the dissipation parameters. We used the results of the analysis done by \citep{Hansen2010} on the distribution of extrasolar planetary systems with a hot Jupiter, in terms of period, eccentricity, and mass. His results correspond to quality factors in the ranges $2 \times 10^6 < Q_p < 2 \times 10^7$ and $ 4 \times 10^6 < Q_s < 10^8$ for the planet and the star, respectively. 

One immediate consequence is that the planet rotation is almost synchronous; depending on the value adopted for $Q_p$, the synchronisation is reached in less than 100 Myr. This upper limit corresponds to a planet in the less dissipative boundary of the interval given by Hansen and to the case in which the planet is assumed to be initially spinning very fast. In the more favourable cases the synchronisation is reached in 10 Myr. 

The other consequence is the almost unchanged eccentricity and semimajor axis of CoRoT-27b during the system's lifetime. Even if parameters lead to a dissipation greater than allowed by the values determined by Hansen, the variation in the eccentricity during the lifetime of the system is less than half its current value. Tidal dissipation is not strong enough to force the circularisation of the orbit (the currently low eccentricity may be primordial). For the semi-major axis, the variation is almost negligible: only a few thousandths of AU in the considered time span.

However, difficulties appear when the stellar rotation is considered. Presently, the star has a slow rotation, and the transfer of the angular momentum of the orbit to the star via the tides raised on the star should make it faster. If so, the star should be even slower in the past. Simulations with the adopted dissipation factors lead to period values that are abnormally high, on time scales shorter than the system's lifetime. This result contradicts with the fast rotations observed in young star clusters, where the periods are always shorter than a few days \citep[see][]{Gallet2013}. The only way to explain the present star period is to admit that the star rotation is not being accelerated, but is being braked. 

\begin{figure}
 \centering
 \includegraphics[width=\columnwidth]{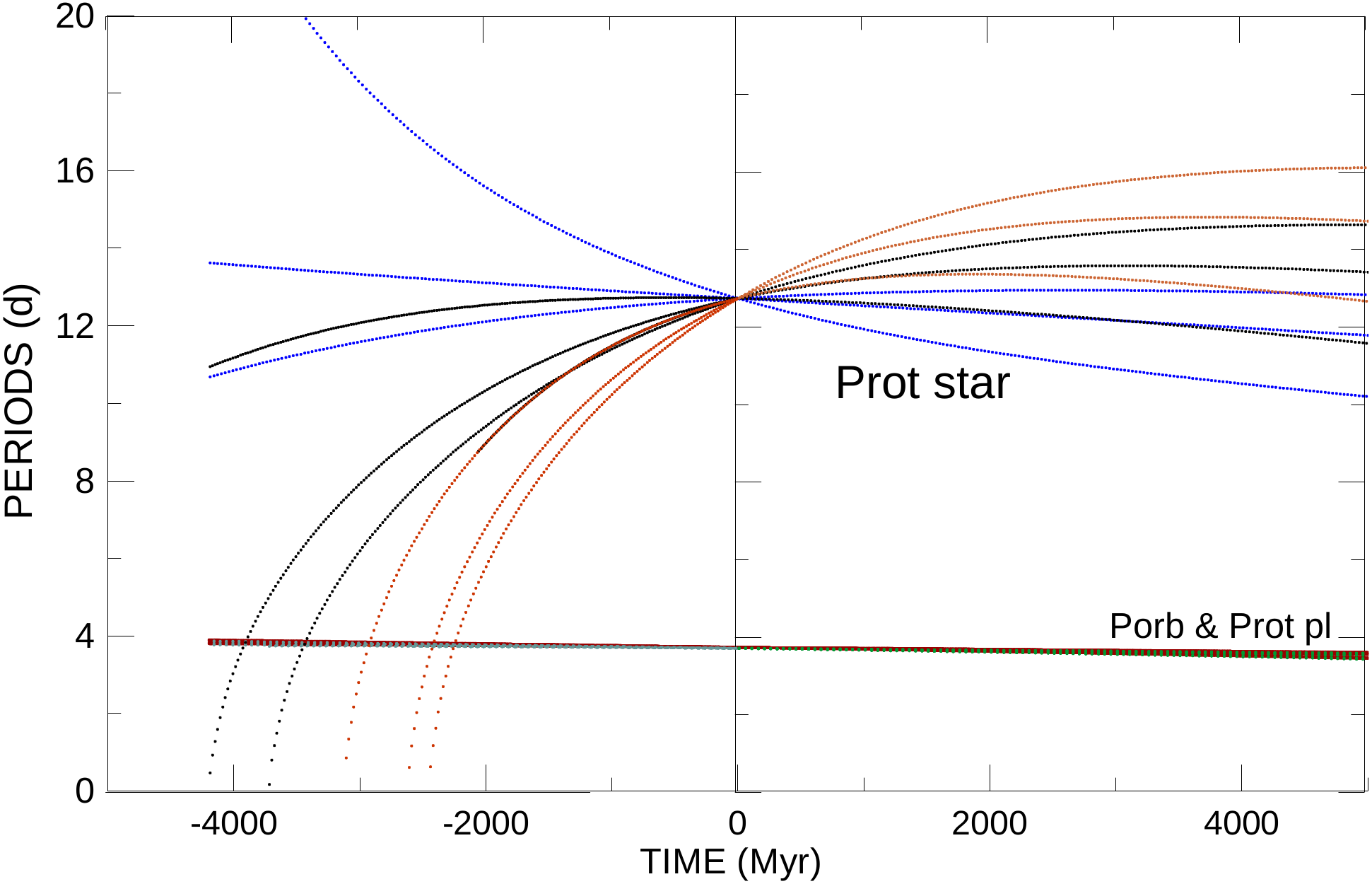}
 \caption{Rotation evolution of the star CoRoT-27 obtained with composite models including magnetic braking, with reduction factors f=0.15~(blue), 0.25~(black) and 0.35~(brown), and the acceleration of the star rotation due to the tidal interaction with the super Jupiter CoRoT 27b, with dissipation values $Q_s = 7 \times 10^6,  10 \times 10^6$, and $13.5 \times 10^6$.}
 \label{fig:tidal_evolution}
\end{figure}

We have studied this possibility with the simple model given by \citet{Bouvier1997} for the angular momentum evolution of low-mass stars due to magnetic braking. However, the calibration constant given by Bouvier et al. is too large and leads to having P=0 in times smaller than the system 
lifetime. It has been conjectured that the given calibration constant needs a correction for stars more massive than the Sun, and \citet{Patzold2012} have corrected it using a factor $f=0.1$ in the study of the tidal evolution of CoRoT-21b around a F-star. A composite model with an undetermined factor $f<1$ and tidal dissipation values consistent with Hansen determinations allows us to find solutions where the initial rotation period is nearly as long as the periods of stars in young clusters and evolves to the presently observed rotation period in ca. 4~Gyr. They are shown in Fig.~\ref{fig:tidal_evolution}. The solutions in this figure correspond to f=0.15, 0.25 and 0.35. We may see that in the worst case (f=0.35), typical dissipation values are not able to avoid a premature reaching of P=0 by the solutions. In each case, three solutions are shown that correspond to dissipation values $Q_s = 7 \times 10^6,  10 \times 10^6$ and $13.5 \times 10^6$ respectively (from up to down in the 
left side of the figure). These values of $Q_s$ refer to the present time. In the two tidal theories used in the modelling \citep{Mignard1979,Ferraz-Mello2013}, the tidal response is fixed by physical properties of the body, and the values of $Q_s$ are not constant but vary with the frequency of the main tide components, that is, with the orbital and rotational periods. 

\subsection{CoRoT-27b in context}
\label{sec:C27_in_context}
CoRoT-27b's mass places it inside the overlapping mass regime between low-mass brown dwarfs and massive planets \citep{Leconte2009,Baraffe2010}. The exact nature of objects in this mass range is not straightforward to establish, and, indeed, depends on the definition of a planet \citep[see][for an overview]{Schneider2011}. 
Definition by mass---whether the object is massive enough to have sustained deuterium fusion at some point of its history---has ambiguities, since the deuterium-burning mass limit can vary from 11 to 16~\mjup depending on the object's metal and helium content~\citep{Spiegel2011}. Also, systems exist with multiple companions likely to be on both sides of the deuterium burning limit \citep{Marcy2001}.
The definition by formation history---whether the object formed by accretion or gravitational collapse---is not without problems either, since we have no reliable means of probing the formation history of an individual object. However, the planet and brown dwarf populations may show some systematic differences on measurable properties, but if such differences exist, more objects are required for any groupings to become discernible.

Considering deuterium burning, CoRoT-27b's $2\sigma$ upper mass limit exceeds the minimum deuterium burning mass limit of 11~\mjup \citep{Spiegel2011}, but is well below the conventional 13~\mjup limit. Thus, it is unlikely, but not completely excluded, that CoRoT-27b would have ever sustained deuterium fusion.

\begin{figure}
 \centering
 \includegraphics[width=\columnwidth]{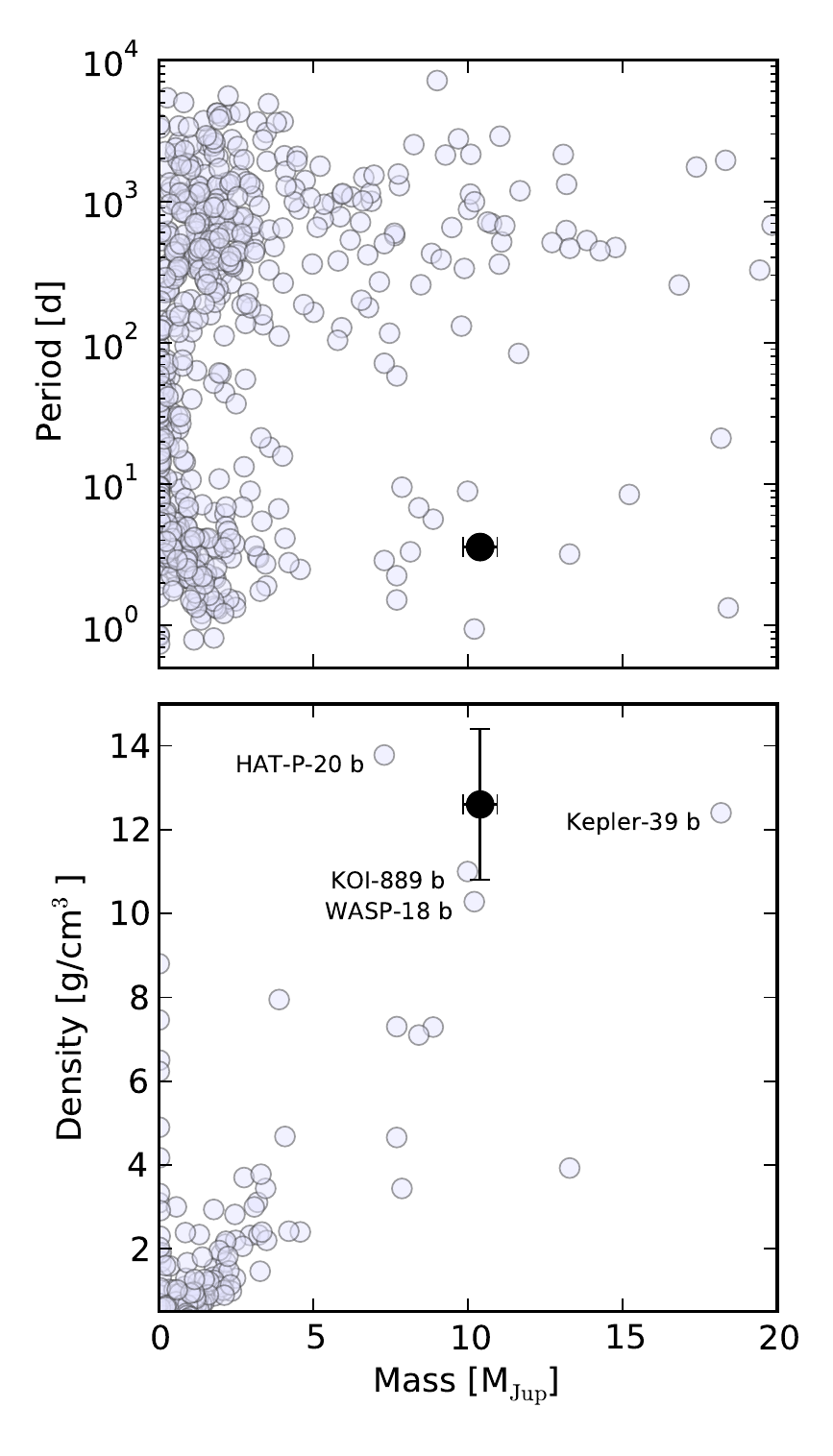}
 \caption{CoRoT-27b mass, period, and density compared with the population of confirmed transiting exoplanets. Planets with masses higher than 20~\mjup and densities higher than 15~\gcm have been excluded.}
 \label{fig:compared_to_other_planets}
\end{figure}

\begin{figure}
 \centering
 \includegraphics[width=\columnwidth]{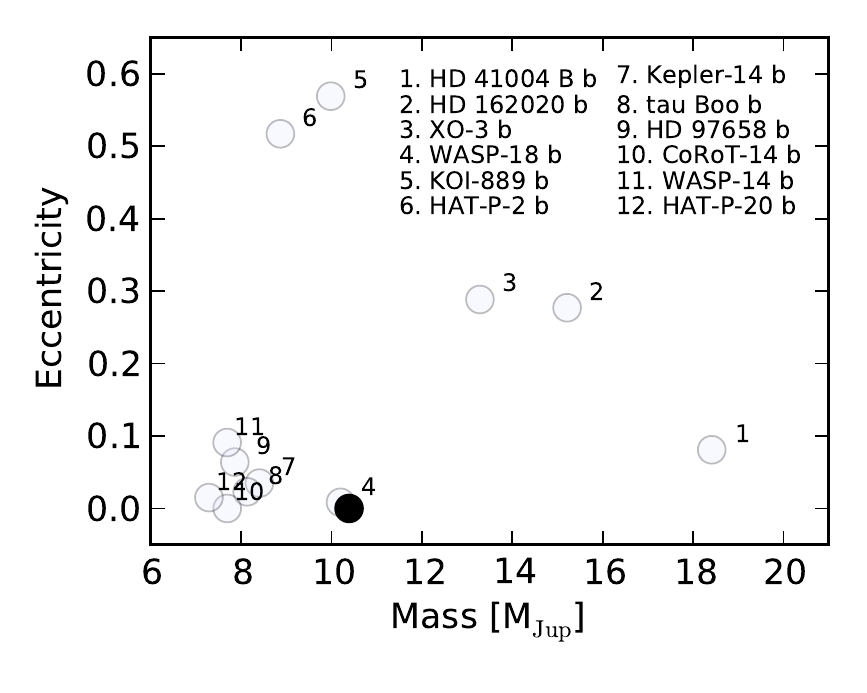}
 \caption{CoRoT-27b mass and eccentricity compared with the population of confirmed transiting exoplanets with periods $<10$~d and masses $5<M_\mathrm{p}<20$~\mjup.}
 \label{fig:mass_and_eccentricity}
\end{figure}

\begin{figure}
 \centering
 \includegraphics[width=\columnwidth]{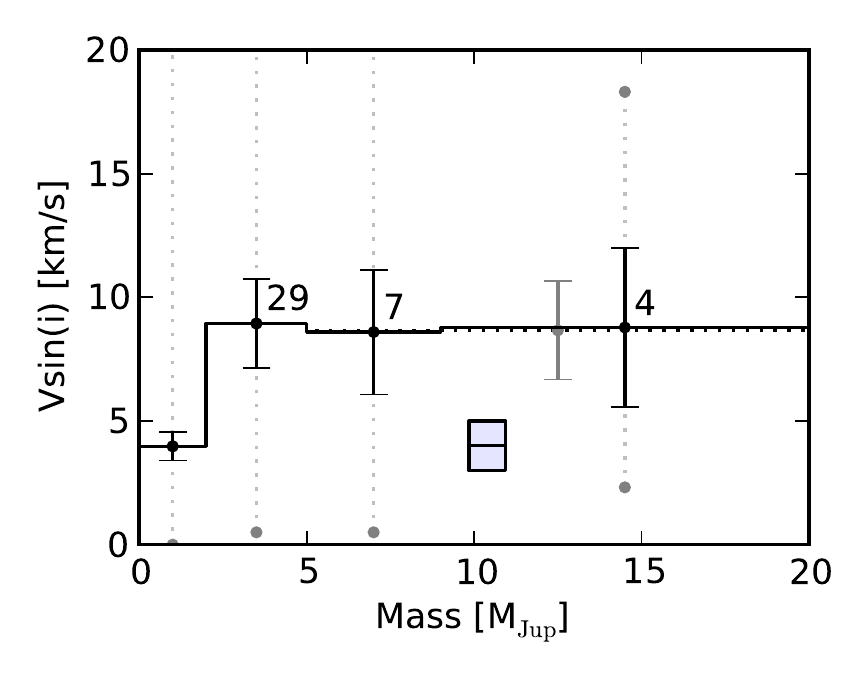}
 \includegraphics[width=\columnwidth]{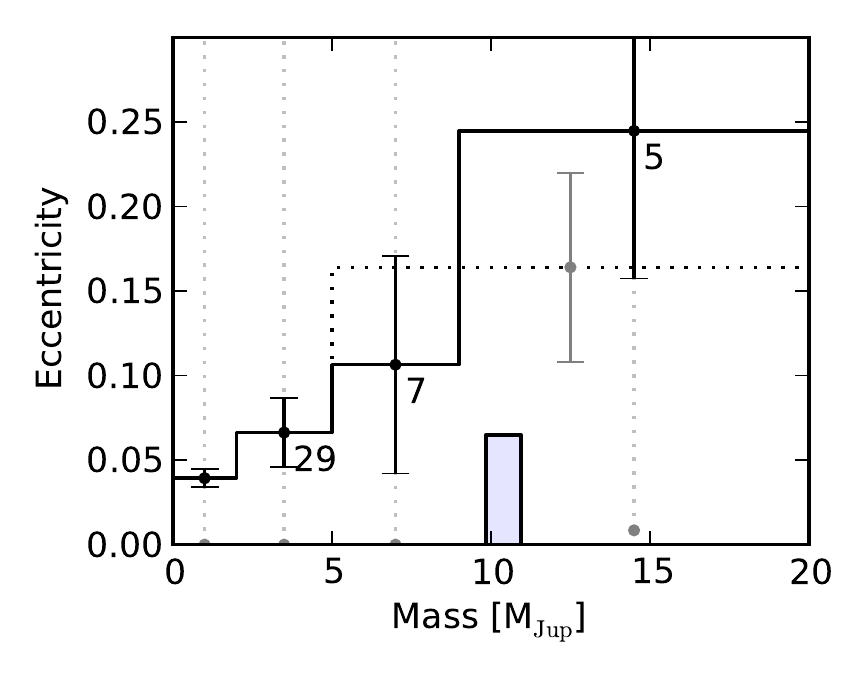}
 \includegraphics[width=\columnwidth]{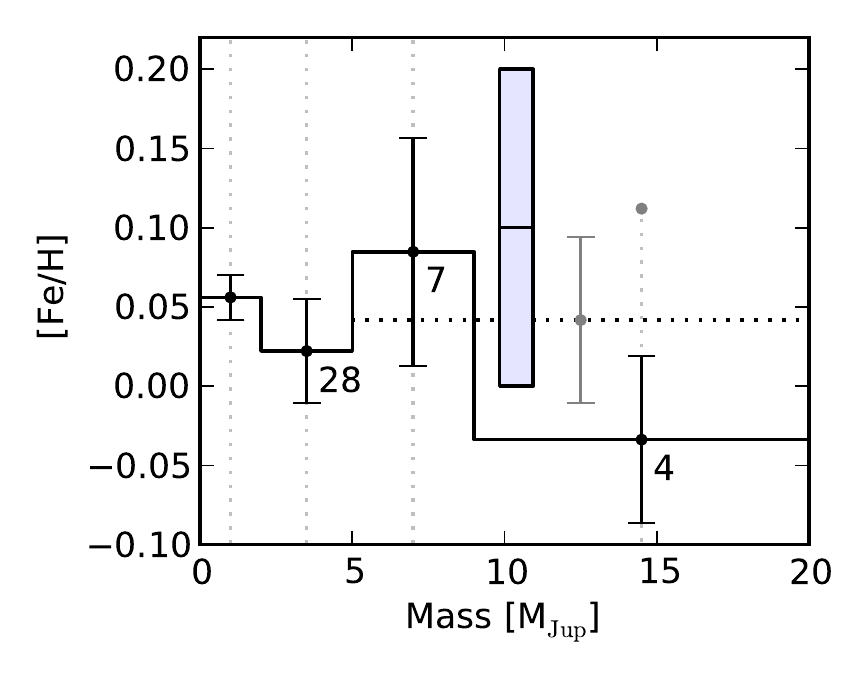}
 \caption{Mean \vsini, orbital eccentricity and host-star metallicity as a function of planet mass for transiting planets with period~$<$10~d. The errorbars correspond to the standard error of the mean, the grey dots show the minima and maxima inside a mass bin, the numbers mark the number of planets in each bin, and the greyed area with a black line shows the parameter estimates with their uncertainty (or the 99\% confidence upper limit at 0.064 for eccentricity). The dotted horizontal line with a grey errorbar shows the average for planets with masses $<5$~\mjup.}
 \label{fig:mass_vs_pars}
\end{figure}

Given the dearth of known massive short-period planets, any statistical analyses are fated to be dominated by small number statistics. Keeping this in mind,
massive short-period planets show a tentative preference to be found orbiting relatively rapidly rotating stars on eccentric orbits \citep[][also Fig.~\ref{fig:mass_vs_pars}]{Bakos2011b,Southworth2009c}, without significant correlation between planetary mass and host-star metallicity \citep{Bakos2011b}. They are also more common around binary systems than single stars \citep{Udry2002}.
We show the CoRoT-27b mass, density, and period compared with the population of transiting exoplanets in Fig.~\ref{fig:compared_to_other_planets}; planetary masses and eccentricities for massive close-in planets in Fig.~\ref{fig:mass_and_eccentricity}; and the average \vsini, eccentricity, and metallicity as a function of the planetary mass in Fig.~\ref{fig:mass_vs_pars}\footnote{From \url{www.exoplanets.org}, accessed 10.8.2013.}. CoRoT-27b can be seen to stand out slightly from the population averages in all cases. While the deviations from the \vsini and metallicity trends are not that significant (inside $2\sigma$ in both cases), the lack of detectable orbital eccentricity is more significant, but not exceptional. All in all, including CoRoT-27b in the population averages of Fig.~\ref{fig:mass_vs_pars} weakens the known trends.

What comes to finding systematic groupings of properties hinting at possible differences in formation and evolution history of objects in the transition region, the currently available set of objects is still too small for any meaningful inferences. Figure~\ref{fig:mass_and_eccentricity} shows two tentative clusters in mass-eccentricity space, with a group of relatively low-eccentricity planets with masses below $11$~\mjup, and another loose group of higher-eccentricity planets. However, no other common factors were identified between the members of two clusters, and many more massive objects are required to confirm (or discard) the significance of these groups.

\section{Conclusions}
We have reported a new massive short-period transiting exoplanet CoRoT-27b, described the properties of the star and the planet, and detailed the analysis methods used to derive them. 

CoRoT-27b was found by the CoRoT-satellite and confirmed by using radial velocities. The stellar properties were determined based on spectral characterisation and further refined with theoretical modelling and transit fitting. We carried out searches for secondary eclipses and additional planets, but found no significant evidence of either from the data.  We described our approach to modelling the planet's structure and composition in Sect.~\ref{sec:structure_and_composition} and concluded that even given its high density, the planet properties can be explained by models with a wide range of heavy element mass fractions. However, the inferred high planetary density may also be a product of underestimated planetary radius due to an unresolved contaminating third light source. Since close-in contaminating sources could not be entirely ruled out, we presented the effects from possible contamination on the planetary radius and density in Sect.~\ref{sec:blending}. We studied the tidal evolution of the system in 
Sect.~\ref{sec:tidal_evolution}, showing that the planet rotation is likely almost synchronous and that the tidal dissipation is not strong enough to strongly affect the orbital eccentricity during the lifetime of the system. Finally, we investigated how CoRoT-27b fits the bigger picture of massive short-period objects within the overlapping mass regime between planets and brown dwarfs. Its properties stand out slightly from the average properties of massive planets, and adding it to the population weakens the tentative trends that have been proposed to separate the massive planets from less massive ones.

All in all, CoRoT-27b is an important addition to a scarcely populated class of massive close-in planets. It is the second object of this type to be found around a G-dwarf, while the rest are predominantly found orbiting hotter F-type stars. More massive short-period planets (and low-mass short-period brown dwarfs) are still required for inferences of any statistical significance, but each new object of this type will help us paint a picture of the differences and similarities between the two populations.

\begin{acknowledgements}
First and foremost we would like to thank the anonymous referee for her/his prompt review.
HP has received support from RoPACS during this research, a Marie Curie Initial
Training Network funded by the European Commission's Seventh Framework Programme.
HP has received funding from the V\"ais\"al\"a Foundation through the Finnish Academy of Science and Letters during this research.
The team at the IAC acknowledges funding by grant AYA2012-39346-C02 of the Spanish Ministry of Economy and Competitiveness (MINECO).
This research was supported by an appointment to the NASA Postdoctoral Program at the Ames Research Center, administered by Oak 
Ridge Associated Universities through a contract with NASA. AS acknowledges the support by the European Research Council/European 
Community under the FP7 through Starting Grant agreement number 239953. The research leading to these results has received funding from the European Union Seventh Framework Programme (FP7/2007-2013) under grant agreement n. 267251.
\end{acknowledgements}

\bibliographystyle{aa}
\bibliography{C27}

\end{document}